\documentclass[11pt,a4paper]{article}
\pdfoutput=1
\usepackage{jheppub}
\usepackage{amsmath,amssymb,cleveref} 
\usepackage{pifont} 
\usepackage{bm}

\usepackage[most]{tcolorbox}

\usepackage[most]{tcolorbox}

\tcbset{
    enhanced,
    breakable,      
    colback=white,
    colframe=black!40,
    boxrule=0.5pt,
    arc=2mm,
    left=6pt,
    right=6pt,
    top=4pt,
    bottom=4pt
}

\newcommand{\lt}{<}
\newcommand{\vp}{\varphi}
\newcommand{\ve}{\varepsilon}

\title{Inflationary Particle Production and the Swampland}

\author[a, b]{Dieter L\"ust}
\author[a]{Joaquin Masias}
\author[c,d]{Mauro Pieroni}
\author[a,e,f]{Marco Scalisi}

\affiliation[a]{Max-Planck-Institut für Physik (Werner-Heisenberg-Institut), Boltzmannstr. 8, 85748 Garching, Germany}

\affiliation[b]{Arnold-Sommerfeld-Center for Theoretical Physics, Ludwig-Maximilians-Universität, 80333, München, Germany}

\affiliation[c]{CERN, Theoretical Physics Department,
Esplanade des Particules 1, Geneva 1211, Switzerland}
\affiliation[d]{
Instituto de Estructura de la Materia (IEM), CSIC, Serrano 121, 28006 Madrid, Spain}

\affiliation[e]{Department of Physics and Astronomy ``Ettore Majorana'', University of Catania, Via S. Sofia 64, 1-95125 Catania, Italy}

\affiliation[f]{INFN-Sezione di Catania, Via Santa Sofia 64, I-95123 Catania, Italy}

\emailAdd{ luest@mpp.mpg.de}
\emailAdd{jmasias@mpp.mpg.de}
\emailAdd{mauro.pieroni@csic.es}
\emailAdd{marco.scalisi@dfa.unict.it}

\abstract{We investigate the impact of particle production during inflation in scenarios where an infinite tower of states features a mass scale that decreases exponentially along the inflationary trajectory. Such couplings naturally arise in string effective field theories and are in fact motivated by the Swampland Distance Conjecture (SDC). We show that the corrections to inflationary observables sourced by the tower scale as $(H/\Lambda_{\text{sp}})^{2+p}$, with $H$ being the Hubble scale, $\Lambda_{\text{sp}}$ being the species scale, that is the quantum gravity cut-off, and $p\geq 1$ characterizes the density of states in the tower. As a result, in gravitationally weakly coupled cosmological effective theories, the tower-induced contributions are suppressed relative to the standard single-field predictions, leaving the inflationary phenomenology essentially unchanged.
We demonstrate this explicitly across a set of well-motivated inflationary potentials, and we compare the resulting predictions with the most recent observational constraints, including those from the Atacama Cosmology Telescope.}

\begin{document}
	\begin{flushright}
		LMU-ASC 25/25\\
		MPP-2025-202
	\end{flushright}
\maketitle

\section{Introduction}
	\label{sec:intro}
Towers of light particles can affect the dynamics of a gravitational system in numerous ways.  One of their universal effects is the renormalization of the scale at which gravity becomes strongly coupled. The particles indeed contribute with a modification of the graviton propagator at 1-loop, and one can show that perturbation theory breaks down at the so-called {\it species scale}~\cite{Dvali:2007hz,Dvali:2007wp,Dvali:2009ks,Dvali:2010vm,Dvali:2012uq} (see also~\cite{Dvali:2001gx,Veneziano:2001ah,Arkani-Hamed:2005zuc,Distler:2005hi} for earlier works), which in four dimensions is
\begin{equation}
		\label{speciescale}
		\Lambda_{\text{sp}} = \frac{M_{\rm P}}{\sqrt{N_{\text{sp}}}}\,,
	\end{equation}
	where $M_{\rm P}$ is the reduced Planck mass and $N_{\text{sp}}$ is the number of particle species with masses below $\Lambda_{\text{sp}}$. At this scale, any effective field theory (EFT) weakly coupled to Einstein gravity breaks down, and quantum gravity features become manifest.
	
	This fact has regained some attention, during the last few years, with the development of the Swampland program \cite{Vafa:2005ui, Ooguri:2006in, Palti:2019pca, Agmon:2022thq}, which aims to identify constraints that quantum gravity sets at the (low energy) effective field theory level. Among these,  the Swampland Distance Conjecture (SDC)~\cite{Ooguri:2006in} is one of the most solid constraints. It states that traversing a super-Planckian distance $\Delta\varphi$ in field space must necessarily be associated with an infinite tower of states having exponentially decreasing mass scale, such as
\begin{equation}
 \label{SDCmass}
	m_{\rm t}\sim~e^{-\gamma\Delta\varphi}\,,
\end{equation}
	with $\gamma$ an $\mathcal{O}(1)$ parameter, whose constraints in realistic setups have been discussed in~\cite{Baume:2016psm, Klaewer:2016kiy,Bedroya:2019snp, Scalisi:2018eaz, Scalisi:2024jhq, Andriot:2020lea,Gendler:2020dfp,Etheredge:2022opl, vandeHeisteeg:2023ubh, Lust:2023zql, Masias:2025hui}. This behavior is typical in string effective models, 
	and according to the Emergent String Conjecture (ESC)~\cite{Lee:2019wij} one can identify the tower with Kaluza-Klein (KK) or string oscillator modes (see also~\cite{Basile:2023blg,Herraez:2024kux}). An important implication of the above formula is the exponential drop-off of the quantum gravity cut-off, as defined in~\cref{speciescale}, namely $\Lambda_{\text{sp}}\sim e^{-\lambda \Delta\varphi}$ with $\lambda$ in general different from $\gamma$, in the same large-distance limit (see \emph{e.g.}~\cite{vandeHeisteeg:2022btw, Castellano:2021mmx}). This means that arbitrarily large super-Planckian field displacements, $\Delta\varphi\gg M_p$, are not allowed in a consistent EFT  as it would necessarily correspond to an arbitrarily small quantum gravity cut-off and, therefore, to a complete breakdown of the theory.
	
\begin{figure}
	    \centering
	    \includegraphics[width=0.55\linewidth]{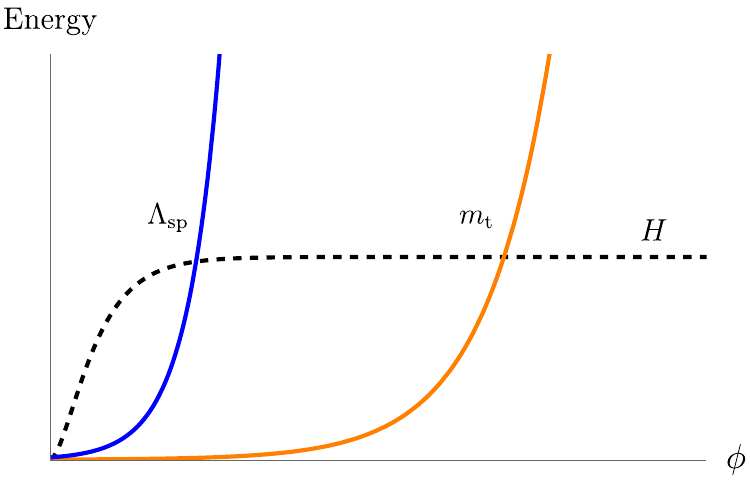}
	    \caption{Exponentially decaying  tower $m_{\text{t}}$ and species $\Lambda_{\text{sp}}$ scales compared to the Hubble scale. There is an intermediate regime in the field excursion of the inflaton in which the inflationary EFT must include the effects of the light tower of species.   }
	    \label{fig:hubblespecies}
	\end{figure}

	In the context of inflationary cosmology, it was shown~\cite{Scalisi:2018eaz} that the requirement that inflationary EFTs have energy scale $H\leq\Lambda_{\text{sp}}$ \cite{Hebecker:2018vxz} leads to a universal upper bound on the inflaton range of the form $\Delta\varphi\lesssim-\log H$ (see also~\cite{vandeHeisteeg:2023uxj,Cribiori:2025oek}). Inflationary scenarios with ranges violating this bound do not admit an effective description weakly coupled to gravity. Here, it is important to notice that the mass scale $m_{\rm t}$ of the tower is parametrically {\it lower} than the cut-off\footnote{One can see this already in the simplest case of an equispaced tower, with $N=\Lambda_{\text{sp}}/m_{\rm t}$ states below the cut-off and masses  $m_n = n\, m_{\rm t}$, with $n\in \mathbb{N}$. Using also~\cref{speciescale}, one obtains that $m_{\rm t} = \Lambda_{\text{sp}}^3\ {M_{\rm P}}^{-2}$, namely the mass scale of the tower is always smaller than the species scale.} $\Lambda_{\text{sp}}$ and it may, therefore, happen that the scale $H$ sits between $m_{\rm t}$ and $\Lambda_{\text{sp}}$. This implies that several states will enter the EFT at energies lower than the gravitational cut-off, with potential observational signatures. In~\Cref{fig:hubblespecies}, we offer a cartoon of this situation for a generic inflationary potential. As one rolls down the potential, the mass of the tower and the UV cutoff decrease exponentially with the field displacement. There is then an intermediate regime, in which one must include the tower in the EFT.

In this article, we show that a mass coupling term such as~\cref{SDCmass}, between the inflaton $\varphi$ and the states of the tower, becomes a source of  {\it particle production} during inflation. This phenomenon has been widely studied in the literature as a potential dissipation channel\footnote{See~\cite{Andriot:2025los,Shiu:2025ycw,Bedroya:2025fwh} for recent work in a similar spirit in the context of dark energy, where (KK) dark matter transfers energy to the quintessence field.}, effectively introducing an additional friction term in the equations of motion. Previous works~\cite{Berera:1998px,Moss:2008lkw, Green:2009ds, Barnaby:2012xt, PhysRevD.80.063533,Pearce:2016qtn, Reece:2022soh} have demonstrated that such effects can be dramatic enough to alter the observational predictions of otherwise uncoupled inflationary models. Recently~\cite{Aoki:2025ywt}, the coupling between the inflaton and additional (heavy) fields has also been explored as a possible mechanism to reconcile inflationary models in tension with the Atacama Cosmology Telescope (ACT) observations~\cite{ACT:2025tim}. Given that a coupling of the form~\cref{SDCmass} arises quite generically in string-theoretic effective field theories, one is naturally led to question whether attempting to embed an inflationary model into a more UV-complete framework might substantially modify its original predictions.\footnote{We pursue this approach by enforcing consistency of the inflationary dynamics with the SDC, without committing to a fully UV-complete, string-theoretic embedding of inflation. In this work, we remain agnostic about the fundamental origin of the potential.}

Our analysis shows that the exponential coupling between the tower and the inflaton, as given in~\cref{SDCmass}, leads to sizable corrections to the sourced power spectrum just in the limit $H\rightarrow\Lambda_{\text{sp}}$, namely when the inflationary model is at the boundary of validity of gravitational weakly coupled effective theories. For a tower of particles with mass spectrum parametrized by $m_n= n^{1/p} m_{\rm t}$, with $p\geq 1$ determining the state density of the tower, we find that corrections to inflationary observables are always given by
\begin{equation}
\delta\{n_s,r,f_{NL}\}\propto \left(\frac{H}{\Lambda_{\text{sp}}}\right)^{2+p}\,,
\end{equation}
where $n_s$, $r$ and $f_{NL}$ are respectively the scalar spectral index, the tensor-to-scalar ratio of primordial fluctuations and the non-gaussianity parameter.

Therefore, the key results of this work are that towers of scalar species with masses given by eq.~\eqref{SDCmass}, as predicted in the context of string theory, produce {\it negligible corrections} to inflationary observables, as long as the model can be described in the regime of weakly coupled EFTs to Einstein gravity.  

Our analysis focuses on a regime away from the strict asymptotic limit in moduli space, where a tower of states already affects the EFT while $H \ll \Lambda_{\rm sp}$ and $V < \Lambda_{\rm sp}^2$. We do not assume slow-roll inflation in the asymptotic regime, where the de-Sitter conjecture \cite{Obied:2018sgi} or the TCC \cite{Bedroya:2019snp} may apply. This is, for example, the regime in which moduli inflation scenarios in string theory (e.g.\ in type IIB model building \cite{Kachrukklt:2003aw,Balasubramanianlvs:2005zx}) are constructed, rather than in the strict asymptotic limit of field space. Although it has been argued that the interior of the string theory moduli space is at most $\mathcal{O}(1)$ in Planck units \cite{vandeHeisteeg:2023uxj,vandeHeisteeg:2022btw,Rudelius:2022gbz,Andriot:2023isc}, we will keep the discussion general, as it is also unclear how directly this extends to general field spaces in the presence of a scalar potential and away from Minkowski limits.

The organization of the article is as follows: In section~\ref{sec:inflation}, we review some aspects of inflation and particle production. Section~\ref{sec:particle_prod_exp} contains our main results and computational details regarding particle production with infinite towers of light modes. In section~\ref{sec:pheno}, we present consequences of our findings to a number of well-motivated inflationary scenarios. Finally, in section~\ref{sec:conclusions}, we conclude.

\section{Slow-roll inflation and inflationary particle production in a nutshell}
\label{sec:inflation}
In its minimal realization, inflation is driven by a single scalar field $\bm{\varphi}$, minimally coupled to gravity and with a canonical kinetic term
\begin{equation}
	S=\int \textrm{d}^4 x \sqrt{-\mathbf{g} }\left[\frac{M_{\rm P}^2}{2}R + \frac{\mathbf{g}^{\mu \nu} }{2} \partial_{\mu} \bm{\varphi} \partial_{\nu} \bm{\varphi} -V(\bm{\varphi}) \right]  \; , 
\end{equation}
where $M_{\rm P} \simeq 2.4 \times 10^{18} $ GeV/$c^2$ is the reduced Planck mass\footnote{In the remainder of this document, we work in natural units, \emph{i.e.}, we set $\bar{h} = c = 1$, and, unless differently stated, we also set $M_{\rm P} = 1$.}, $R$ is the Ricci scalar, and $V( \bm{\varphi})$ is the scalar field potential. The system is then expanded in terms of a homogeneous background plus perturbations as
\begin{equation}
	\label{eq:perturbative_expansion}
	\bm{\varphi}(t, \vec{x}) =\varphi(t) + \delta\varphi(t, \vec{x}) \; ,  \qquad \mathbf{g}_{\mu\nu}(t, \vec{x}) = g_{\mu\nu}(t) + \delta g_{\mu\nu}(t, \vec{x}) \; ,
\end{equation}
where $\vp(t)$, $g_{\mu\nu}(t)$ denote the homogeneous background fields and $\delta\varphi(t, \vec{x})$, $ \delta g_{\mu\nu}(t, \vec{x})$ denote the perturbations around the background. In cosmic time $t$, the action for the background inflaton field reads
\begin{equation}
	S_{\varphi} = \int \textrm{d}^4 x\ a^3 (t) \left[ \frac{\dot{\varphi}^2}{2 } -V(\vp) \right]  \; , 
\end{equation}
where we have substituted a Friedmann–Lema\^{i}tre–Robertson–Walker (FLWR) metric, with a mostly negative signature, with $a(t)$ representing the scale factor, and a dot denoting a derivative with respect to $t$. The system is fully characterized by the scalar field equation of motion (e.o.m.)
\begin{equation}
	\ddot{\varphi} + 3 H \dot{\varphi} + V_{, \varphi}  = 0 \; , 
\end{equation} 
where $V_{, \varphi} \equiv \partial V / \partial \varphi $, together with the only two independent components of the Einstein equations
\begin{equation}
	3 H^2  =  \frac{\dot{\varphi}^2}{2 } + V(\vp) \; , \qquad \qquad  - 2 \dot{H} = \dot{\varphi}^2 \; , \label{eq:EE_inflation} 
\end{equation}
where $H \equiv \dot{a} / a$ is the Hubble parameter. Inflation corresponds to a phase of quasi de-Sitter $a(t) \simeq e^{Ht}$ expansion of the Universe, requiring $\dot{H} \ll H^2$. From~\cref{eq:EE_inflation}, it is clear that this implies $\dot{\varphi}^2 \ll V$. For this reason, this scenario is typically dubbed slow-roll inflation, and for its characterization, it is convenient to introduce the so-called slow-roll parameters (see, \emph{e.g.},~\cite{Baumann:2009ds})
\begin{equation}
	\ve \equiv - \frac{\dot{H}}{H^2} = \frac{\textrm{d} \ln H}{\textrm{d} N_e} =  \frac{1}{2} \frac{\dot \varphi^2}{H^2} \simeq \frac{1}{2} \left(\frac{\textrm{d}\ln V}{\textrm{d}  \varphi} \right)^2 \; , 
	\label{eq:epsilon}
\end{equation}
\begin{equation}
\label{eq:eta}
\eta \equiv -\frac{\ddot \varphi}{\dot \varphi H } = \frac{\textrm{d} \ln \dot \varphi}{\textrm{d} N_e}  =  \frac{1}{2} \frac{\textrm{d} \ln \ve }{\textrm{d} N_e} + \ve \simeq \frac{V_{\varphi \varphi}}{V} - \ve\; .
\end{equation}
From these definitions, the condition $\ve \ll 1 $ ensures that the Universe experiences accelerated expansion, while $\eta \ll 1 $ imposes that this configuration is preserved for enough time. Finally, we define the number of e-folds $N_e $ as
 \begin{equation}
     \textrm{d} N_e \equiv -H \textrm{d}t \;,
     \label{eq:N_definition}
 \end{equation}
 which provides a convenient characterization of time during inflation. In the following, we work using the convention giving $N_e =0$ at the end of inflation and large and positive $N_e$ deep in the inflationary phase.

Typically, the behavior of perturbations is studied by performing a scalar, vector, tensor decomposition of $\delta\varphi(\tau, \vec{x})$ and $\delta g(\tau, \vec{x})$, and by identifying some gauge invariant combinations (for a review, see, \emph{e.g.},~\cite{Mukhanov:1990me}). At the end of this procedure, the system will depend on only two physical quantities\footnote{Vector perturbations do not contribute to the formation of large scale structure in the early universe and, if produced, would nonetheless decay rapidly~\cite{Kodama:1984ziu,MUKHANOV1992203}.} (for a total of three degrees of freedom): the comoving curvature perturbation $\mathcal{R}$, and the traverse traceless spatial tensor $h_{ij}$, consisting of two degrees of freedom corresponding to the two gravitational wave (GW) polarizations. By studying the equations of motion for these quantities, one can show that perturbations generated on sub-horizon scales are stretched by the expansion until they exceed the Hubble radius at that time, at which point they effectively freeze. At later times, when these perturbations become causal again, they leave an imprint on the Cosmic Microwave Background (CMB) temperature and polarization~\cite{MUKHANOV1992203, Seljak:1996ti, Seljak:1996gy}. Among the quantities CMB observations can probe, there are the (dimensionless) scalar and tensor power spectra computed at horizon crossing
	\begin{equation}
		P_{\zeta}=\frac{k^3}{2\pi^2} \, \dfrac{H^2}{\dot\vp_0^2 }\left|\delta\vp \right|^2 = \frac{H^2}{8 \pi^2 \varepsilon }\, , \qquad  P_{T}=\frac{k^3}{2\pi^2} \, 4 \sum_\lambda \left|h_\lambda \right|^2 = \frac{2 H^2}{\pi^2  }\,\, ,
	\label{eq:power_spectra_vacuum}
	\end{equation}
	where $\lambda = +, \times$ denote the two GW polarizations. In these expressions, we have used the slow-roll parameters defined in~\cref{eq:epsilon,eq:eta}, and we have substituted the solutions for $\delta \varphi$ and $h_{\lambda}$. The constraints set by CMB observations are typically expressed in terms of the scalar spectral index $n_s$, which measures the departure from scale invariance, and the tensor-to-scalar ratio $r$, which  measures the intensity of tensor perturbations compared to their scalar counterpart
	\begin{equation}
    \label{eq:nsr00}
		n_s  = 1 +  \frac{\textrm{d} \ln \, {P}_\zeta }{ \textrm{d} \ln k} \simeq  1 + 2 \eta - 4 \varepsilon   \, , \qquad    r= \dfrac{P_T}{P_\zeta} =  16 \varepsilon \, ,
	\end{equation} 
	where in the expression of $n_s$ we have kept only the lowest order terms in slow-roll parameters. The most recent experimental bounds on the amplitude of the scalar power spectrum, $n_s$ and $r$ read~\cite{Planck:2018vyg, Planck:2018jri,BICEP:2021xfz}
	\begin{equation}
		\left. P_{\zeta} \right|_{k_*} = (2.10 \pm 0.03) \times 10^{-9} \; , \qquad \left. n_s-1\right|_{k_*}=0.035 \pm 0.004 \; , \qquad \left. r\right|_{k_*}<0.036 \; ,
		\label{eq:ns_r_bounds}
	\end{equation}
    where $k_* = 0.002 {\rm Mpc}^{-1}$ denotes the pivot scale at which these quantities are measured. The first of these constraints, which sets the amplitude of the scalar power spectrum at CMB scales, is also referred to as the COBE normalization. For slow-roll inflation, $k_*$ can be easily connected to the number of e-folds to the end of inflation (see, \emph{e.g.},~\cite{Liddle:2003as, Martin:2010kz}). So that, in practice, given an inflationary model, it is sufficient to evaluate the right-hand sides of~\cref{eq:power_spectra_vacuum,eq:nsr00} at $50 \lesssim N_e \lesssim 60$ and compare with the constraints in~\cref{eq:ns_r_bounds}.

	While the power spectra are proportional to the two-point correlators of $\mathcal{R}$ and $h$, similar quantities can be associated with higher-order correlation functions. In particular, the scalar bispectrum $B_{\zeta}$ is obtained by computing cubic correlations. In most single-field inflation models, the bispectrum is typically strongly suppressed at large scales, implying that primordial perturbations are highly Gaussian. For more general models, non-gaussianities can be quantified in terms of the non-linearity parameter defined as 
 
	\begin{equation}
		\label{eq:FNL_def}
		f_{\rm NL}=B_\zeta (k_1,k_2,k_3)\dfrac{10}{3}\dfrac{1}{(2\pi)^{5/2}}\dfrac{}{P_\zeta^2}\dfrac{\prod_i k_i^3}{\sum_i k_i^3} \; ,
	\end{equation}
	where $B_\zeta$ is the scalar bispectrum
	\begin{equation}
		\label{eq:bispectrum_def}
		B_\zeta (\vec{k}_1,\vec{k}_2,\vec{k}_3)=-\dfrac{H^3}{\dot{\vp_0}^3}a^
		{-3}\langle \delta\vp (\tau, \vec{k}_1)\text{ } \delta\vp (\tau, \vec{k}_2)\text{ } \delta\vp (\tau, \vec{k}_3) \rangle \; .
	\end{equation}
    In the case of particle production~\cite{Pearce:2016qtn, PhysRevD.80.063533, Barnaby:2010vf, Barnaby:2011qe, Barnaby:2012xt}, the spectrum is expected to peak for equilateral configurations, corresponding to $k_1=k_2=k_3$. The current bound on equilateral non-gaussianities (68$\%$ CL) is given by~\cite{Planck:2019kim}
	\begin{equation}
		f_{\rm NL, equil}=-26\pm 47 \; , 
	\end{equation}
	implying that non-gaussianities are compatible with zero.

	\subsection{Basics of inflationary particle production}
	\label{sec:particle_production_review}

    In realistic scenarios, the inflaton might interact with other degrees of freedom. Such interactions can lead to particle production during inflation, providing an additional channel for energy transfer that modifies the background evolution and the dynamics of perturbations. In particular, light or weakly massive fields can be produced efficiently, sourcing curvature and tensor fluctuations beyond the standard single-field predictions. These effects may reduce the tensor-to-scalar ratio and potentially bring otherwise excluded models, such as simple monomial potentials, back into agreement with current observations~\cite{PhysRevD.80.063533, Pearce:2016qtn, Cook:2011hg, Dufaux:2007pt, Planck:2019kim, BICEP:2021xfz}. In the following, we review the basics of particle production during inflation through one of its simplest realizations.
	
	A natural way to couple the homogeneous part of the inflaton $\vp$ to an additional (inhomogeneous) field $\chi$ is via a quadratic coupling of the form
	\begin{equation}
		S = \int \mathrm{d}^{4} x \sqrt{-g}\left[\frac{g^{\mu \nu} }{2} \partial_{\mu} \vp \partial_{\nu} \vp-V(\vp)+\frac{g^{\mu \nu} }{2} \partial_{\mu} \chi \partial_{\nu} \chi-\dfrac{1}{2}f(\vp)\chi^2\right]  \;  ,
	\end{equation}
	where $\sqrt{f(\vp)}$ represents a $\vp$-dependent mass for the field $\chi$. Given an unperturbed FLRW metric 
 \begin{equation}
     \mathrm{d}s^2=g_{\mu\nu}\mathrm{d}x^{\mu}\mathrm{d}x^{\nu}=\mathrm{d}t^2-a(t)^2\delta_{ij}\mathrm{d}x^{i}\mathrm{d}x^{j},
 \end{equation}
 the e.o.m. for the (homogeneous) inflaton and the (inhomogeneous) $\chi$ field read
	\begin{align}
		\label{eq:varphi_eom_mod}
		\ddot\vp+3 H \dot\vp +V_{, \varphi} + \frac{1}{2} \, f_{,\vp}(\vp)\langle\chi\chi\rangle
		& = 0 \; , \\
		\label{eq:chi_eom2}
		\ddot{\chi}+3 H \dot{\chi}-\frac{\nabla^{2}\chi} {{a^{2}}}+f(\vp)  \chi &= 0  \;  ,
	\end{align}
 where the brackets denote a space average.\footnote{Notice that only space averaged quantities built out of the $\chi$ field can contribute to the e.o.m. for the homogeneous inflation.} The term in~\cref{eq:varphi_eom_mod}, proportional to $\langle\chi\chi\rangle$, represents the $\chi$ backreaction on the inflaton's trajectory. We stress that here (and throughout the rest of the paper) we assume no (homogeneous) zero mode to be present for the $\chi$ field(s). Namely, we assume $\langle \chi \rangle = 0$, so that only the variance or higher order correlation functions of the $\chi$ field(s) can contribute to the dynamics. The presence of a homogenous contribution from the $\chi$ field(s) would otherwise further affect the background dynamics. 
 
 It is convenient to express the field equation for $\chi$ in terms of conformal time $\tau$, defined by $\textrm{d}\tau=\, \textrm{d}t/a$, as well as to re-scale the field as $\chi=\xi/a$. This allows for a simpler e.o.m. as
\begin{equation}
		\xi''-\nabla^2\xi + a^2 f(\vp)\xi -\dfrac{a''}{a}\xi = 0 \; , \label{eq:chieq}
	\end{equation}
	where  $'$ denotes a derivative with respect to conformal time. 
 We can express this in terms of momentum modes
 \begin{equation}
     \xi(\tau,\vec{k})=\int\dfrac{\textrm{d}^3x}{(2\pi)^{3/2}} \; e^{-i\vec{k}\cdot\vec{x}} \; \xi(\tau,\vec{x}) \; ,
 \end{equation}
and the  e.o.m. for the $\xi_n$ field can be written as a wave equation of the form
	\begin{equation}
 \label{eq:introeqchi}
		\xi''(\tau, \vec{k})+\omega^2 \, \xi(\tau, \vec{k})=0\; .
	\end{equation}
	where we have introduced
	\begin{equation}
		\omega^2=\left[k^2 + a^2f(\vp)-\dfrac{2}{\tau^2}\right] \; .
	\end{equation}

When the effective mass $a^2f(\vp)$ becomes small then $\omega^2$ can easily approach zero, thus making the production of $\xi$ modes kinematically favored. A negative $\omega^2$ induces instead a tachyonic instability, enhancing the amplitude of the
$\xi$ modes. Both regimes can be realized through a suitable choice of the coupling function $f(\vp) $. In particular, one can engineer this behavior to occur at one~\cite{Barnaby:2009mc,Kofman:1997yn,Kofman:2004yc} or more~\cite{Green:2009ds,Pearce:2016qtn} points along the inflaton trajectory. Coupling the inflaton to an infinite tower of states has been thoroughly explored in the context of Trapped Inflation~\cite{Green:2009ds}. In such a scenario, particle production occurs continuously, causing significant dissipation and thereby slowing the inflaton’s motion. Additionally, the scalar power spectrum is enhanced due to the excitation of the extra degrees of freedom, which suppresses the tensor-to-scalar ratio and can induce substantial modifications to the spectral tilt~\cite{Pearce:2016qtn, PhysRevD.80.063533}.

\section{Particle production from a tower with  an exponential mass scale}
\label{sec:particle_prod_exp}	
Let us now consider a model where the inflaton $\bm{\varphi}$ is coupled to an infinite tower of states $\chi_n$, with masses exponentially sensitive to the inflaton value\footnote{The precise behaviour of the mass scale of the tower (whether it grows or decreases during inflation) depends on $\gamma$ and on the dynamics of $\bm{\varphi}$. In the following, we cover both scenarios by considering both positive and negative values for $\gamma$ and comment on this point where relevant.}
	\begin{equation}
		S=\int \textrm{d}x^4 \sqrt{-\mathbf{g}}\left[\frac{\mathbf{g}^{\mu \nu} }{2} \partial_{\mu} \bm{\varphi}\partial_{\nu} \bm{\varphi}-V(\bm{\varphi})+\sum_n \left( \dfrac{\mathbf{g}^{\mu \nu} }{2} \partial_\mu\chi_n\partial_\nu\chi_n- \dfrac{m_n^2}{2} e^{-2\gamma \bm{\varphi} } \chi_n^2 \right) \right] \;,
    \label{eq:full_action_exp}
	\end{equation}
    with
    \begin{equation}
	   m_n=n^{1/p} m_1\,,
        \label{eq:scalingofmasses}
	\end{equation}
where $n$ is a positive integer labeling the modes, $p$ is a scaling exponent that determines how the mass grows with the level $n$ and $m_1$ is the mass scale $m_n$ for the first state of the tower.
    
As in~\cref{sec:inflation}, we expand the fields $\bm{\varphi}$ and $\mathbf{g}_{\mu \nu}$ in terms of a homogeneous background plus perturbations. Substituting into~\cref{eq:full_action_exp}, we get the e.o.m. for the homogeneous background, which in terms of the physical time $t$, reads
    \begin{align}
    \label{eq:eom_varphi_backreactions}
		\ddot\varphi +3 H \dot\varphi + V_{, \varphi}- \sum_n \gamma \, m_n^2  e^{-2\gamma \varphi } \langle\chi_n^2 \rangle =0\; , \\ \label{eq:chi_eom}
		\ddot{\chi}_n+3 H \dot{\chi}_n-\frac{\nabla^{2}\chi_n} {{a^{2}}}+m_n^2 e^{-2\gamma\varphi } \chi_n &= 0  \;  ,
	\end{align}
	On the other hand, Friedmann's equation reads
	\begin{equation}
		\label{eq:H_backreactions}
		3 H^2   = \frac{\dot{\varphi}^2}{2} + V(\vp) + \sum_n \left[ \frac{\langle \dot{\chi_n}^2 \rangle }{2} + \dfrac{m_n^2}{2} e^{-2\gamma \varphi } \langle \chi_n^2 \rangle \right]  \; .
	\end{equation}
	Similarly, we can compute the e.o.m. for $\delta\varphi$, which also depends on $\langle \chi_n^2 \rangle$ (see~\cref{sec:perturbations}). It is clear from~\cref{eq:eom_varphi_backreactions} that assessing the impact of particle production on the inflaton dynamics and the inflationary observables requires evaluating the two-point functions of the fields $\chi_n$. 

 As in~\cref{sec:particle_production_review}, we introduce 
 \begin{equation}
     \xi_n \equiv a(\tau) \chi_n \; ,
 \end{equation}
 so that the e.o.m. for the $\xi_n$ fields reads
	\begin{equation}
		\xi''_n(\tau, \vec{k})+\omega_n^2 \, \xi_n(\tau, \vec{k})=0\; ,
		\label{eq:chieq2}
	\end{equation}
	where $\omega_n$ is given by
	\begin{equation}
		\label{eq:omega_n}
		\omega_n^2=\left[k^2 + a^2m_n^2 e^{-2\gamma \varphi}-\dfrac{2}{\tau^2}\right] \; .
	\end{equation}
As discussed in~\cref{sec:particle_production_review},	negative values of $\omega_n^2$ introduce a tachyonic instability in the theory, leading to the explosive production of the corresponding modes $\xi_n$. Before studying in detail the conditions for such a phenomenon to occur in the model of our interest, we discuss the constraints to which this scenario is subject.
	
	\subsection{Quantum gravity constraints on the model}
	\label{sec:constraints}

    The presence of a tower of infinite scalars necessarily renormalizes the scale at which quantum gravity (QG) effects become relevant. This gravitational cut-off is the so-called species scale, as given in	eq.~\eqref{speciescale}. For the scenario considered in this work, the characteristic mass of the tower of states is given by
	\begin{equation}\label{eq:mt}
		m_{\text{t}}=m_1\ e^{-\gamma \vp/M_{\rm P}}\; ,
	\end{equation}
    where we assume that the reference scale for the tower depends on the v.e.v. of some moduli $\phi_i$ in the UV theory, that is $m_1=m_1(\phi_i)$.\footnote{Additionally, we assume that the masses are protected against thermal corrections, for instance through softly broken supersymmetry.} A similar assumption was made in~\cite{Anchordoqui:2025epz}, where the mass scale of the tower was fixed by the size of the internal manifold, while the dilaton rolled down the top of the scalar potential. The exponential scaling $\gamma$ is bounded~\cite{Etheredge:2022opl, Lee:2019wij} in $D=4$ such as
    \begin{equation}
        \sqrt{\frac{1}{2}}\leq|\gamma|\leq\sqrt{\frac{3}{2}}\,
        \label{eq:sdconst}
    \end{equation}
    where the upper bound corresponds to a decompactification limit from $D=4$ to $D=5$, while the lower bound is valid for limits with weakly coupled critical strings.
 
	Here and in the following sections we focus on $p=1$, see eq.~\eqref{eq:full_action_exp}, which corresponds to an equi-spaced tower. We will later present the generalization to arbitrary $p$ in Appendix~\ref{sec:appendix_powers}. Identifying the modes $\chi_n$ with states of a KK tower, the case $p=1$ amounts to having only a single compact dimension becoming large. For $p=1$, the mass of the heaviest state below the cutoff is given by $N_{\text{sp}}\,m_{\text{t}}$, so we have the relation
	\begin{equation}
		N_{\text{sp}}\,m_{\text{t}}\simeq \Lambda_{\text{sp}}  \; .
	\end{equation}
	 This allows us to express the species scale as
	\begin{equation}
 \label{eq:lambda_species}
		\Lambda_{\text{sp}}=M_{\rm P}^{2/3}m_{\text{t}}^{1/3}=M_{\rm P} \left(\dfrac{m_1}{M_{\rm P}}\right)^{1/3}e^{-\gamma \varphi/3 M_{\rm P}}  \;  , 
	\end{equation}
	and the number of modes below the species scale as
	\begin{equation}
		N_{\text{sp}} =\left(\dfrac{M_{\rm P}}{m_{\text{t}}}\right)^{2/3}=\left(\dfrac{M_{\rm P}}{m_1}\right)^{2/3}e^{2\gamma\vp/3M_{\rm P}}  \;  .
	\end{equation}
	We observe that, along a field excursion, the species scale decreases exponentially, while the number of species grows exponentially. Note that the scalings of $\Lambda_{\text{sp}}$ and $N_{\text{sp}}$ differ from the scaling of the mass $m_{\rm t}$ as given above (see \cite{Castellano:2023stg,Basile:2025bql} for a recently uncovered universal pattern relating these scalings).
    
    Given the above considerations, the model described by the action in eq.~\eqref{eq:full_action_exp} must obey at least two constraints.

\begin{tcolorbox}[title={ Constraint C1}: Energy scale]
The inflationary energy scale must lie well below the quantum gravity cutoff to ensure a consistent, weakly-coupled gravitational EFT \cite{Hebecker:2018vxz,Scalisi:2018eaz}. This amounts to imposing
	\begin{equation}
		   H  \ll \Lambda_{\text{sp}} \; .
		\label{eq:expca2}
	\end{equation}
	This is equivalent to requiring that the UV cutoff is higher than the IR scale, which in quasi-de Sitter space is precisely given by $H$. We now introduce $N_H$ as the number of species below the Hubble scale
	\begin{equation}
		N_H= \dfrac{H}{m_{\rm t} }=\dfrac{H}{m_1 } e^{\gamma \vp/M_{\rm P}}  \;  ,
		\label{eq:NH}
	\end{equation}
    that relates to the species scale as 
    \begin{equation}
    \label{eq:NH_Lambda_eq}
        N_H \left( \frac{H}{M_{\rm P}}\right)^2 = \left(  \frac{H}{\Lambda_{\text{sp}}} \right)^3 \; .
    \end{equation}
	Then, we can express the constraint~\cref{eq:expca2} as
	\begin{equation}
		N_H \ll  N_{\text{sp}}\;.
		\label{eq:expca22}
	\end{equation}
	We notice that both quantities, $N_H$ and $N_{\text{sp}}$, are bounded from above by $H^{-2}$ (in Planck units), as expressed by
\begin{equation}
    N_H \ll N_{\rm sp} = \frac{M_{\rm P}^2}{\Lambda_{\text{sp}}^2} \ll \frac{M_{\rm P}^2}{H^2} \; .
    \label{eq:expca222}
\end{equation}

This hierarchy admits a simple geometric interpretation. Each light species below the quantum gravity cutoff can be thought of as occupying a Planck-area element $l_{\rm P}^2 = 1/M_{\rm P}^2$. The total number of species $N_{\rm sp}$ must then fit within the area of the Hubble patch, $R_H^2 \sim 1/H^2$, thus yielding
\begin{equation}
N_{\rm sp} \, l_{\rm P}^2 < R_H^2 \,.
\end{equation}
The notable aspect of this relation is that the relevant quantity controlling the bound is the \emph{area} of the Hubble horizon, rather than its volume. This mirrors what happens for black holes, where the number of independent degrees of freedom that can be consistently stored in a region is also set by an area law. The fact that de Sitter space exhibits the same type of areal constraint is therefore non-trivial, and suggests that species bounds are deeply connected to horizon thermodynamics rather than to local field-theoretic counting (see e.g. \cite{Cribiori:2023ffn,Herraez:2024kux,Herraez:2025clp}).

\end{tcolorbox}

\begin{tcolorbox}[title=Constraint C2: Time scale]

The species scale $\Lambda_{\text{sp}}$ varies exponentially with $\varphi$ as given in eq.~\eqref{eq:lambda_species}. Therefore, to avoid violating the constraint C1 in eq.~\eqref{eq:expca2} along the inflationary trajectory, we must require that the characteristic timescale $t_{\rm sp}$ over which the species scale changes be much longer than the Hubble timescale $t_H$ during inflation., namely
\begin{equation}\label{constraintspeciesH}
		 t_{\rm sp}\ll t_H\; .
	\end{equation}

 The time scale of quantum gravity effects can be deduced from  eq.~\eqref{eq:lambda_species} by considering $\varphi\simeq \varphi_0+\dot\varphi, t$, which gives
	\begin{equation}
    \label{eq:Lambda_QG_label}
		\Lambda_{\text{sp}}\propto e^{-\gamma\dot\vp\,t/3M_{\rm P}} \qquad \longrightarrow  \qquad  t_{\rm sp} \equiv \dfrac{3M_{\rm P}}{|\gamma\dot\vp|}\; .
	\end{equation}
The timescale of inflation is read off from the evolution of the scale factor,
	\begin{equation}
		a(t)\simeq e^{H t} \qquad \longrightarrow  \qquad  t_H \equiv \dfrac{1}{H}\; .
	\end{equation}
	
	Imposing the condition in eq.~\eqref{constraintspeciesH} then leads to
	\begin{equation}
		\dfrac{|\gamma\dot\vp|}{3 H M_{\rm P}}\ll 1\; .
		\label{eq:expca1}
	\end{equation}
	which is a physical requirement of the model, ensuring that inflation occurs before the breakdown of the EFT. Equivalently, this condition guarantees that the masses in the tower vary adiabatically on inflationary timescales, \emph{i.e.},
\begin{eqnarray}
    \dfrac{\dot m_t}{m_t} \, t_H \ll 1.
\end{eqnarray}

\end{tcolorbox}

	\subsection{Particle production}
	\label{sec:particle_production}
	In this section, we analyze the instability of the $\chi_n$ fields and its impact on the background evolution, under the constraints introduced in~\cref{sec:constraints}. Approximating the evolution as quasi–de Sitter with $a \simeq - 1 / (H\tau)$ we can rewrite~\cref{eq:omega_n} as
	\begin{equation}
		\omega_n^2=\left[k^2 + \dfrac{m_n^2}{H^2\tau^2}  e^{ -2\gamma \varphi } -\dfrac{2}{\tau^2}\right]
		\label{eq:omegaii}\, .
	\end{equation}
	We proceed by introducing $\delta_n$, as the squared ratio between the mass of the n$^\text{th}$ mode in the tower and the Hubble scale 
	\begin{equation}
		\label{eq:delta_def}
		\delta_n \equiv \dfrac{m_n^2}{H^2}\ e^{-2\gamma \varphi }=\frac{(n\ m_t)^2}{H^2} \, \; 
	\end{equation}
   where, in the last equality, we have used eqs.~\eqref{eq:scalingofmasses} and \eqref{eq:mt}, and set $p=1$, as we will assume throughout the rest of the analysis. In terms of this quantity,~\cref{eq:chieq2} reads
	\begin{equation} 
    \xi''_n(\tau, \vec{k})+\left[k^2 -\dfrac{2-\delta_n}{\tau^2}\right]\xi_n(\tau, \vec{k})=0\; . \label{eq:xi_eq_2}
	\end{equation}
    Assuming $\delta_n$ to be nearly constant, which is valid under the slow-roll approximation, the solution of this equation is given in terms of the Bessel function of the third kind (typically referred to as the Hankel function of the first kind) $H^{(1)}$ as
	\begin{equation}
		\xi_n(\tau,\vec{k})=\frac{\sqrt{- \pi }}{2} \, \exp\left[\frac{i \pi}{4}   \sqrt{9-4 \delta_n }+\frac{i \pi }{4}\right] \sqrt{-\tau} \, H_{\frac{1}{2}\sqrt{9-4\delta_n} }^{(1)}(-k \tau) \; .
		\label{eq:xisol1}
	\end{equation}
	
	This solution features two different regimes. At early times, corresponding to $|k\tau|\gg 1$, the modes are well inside the horizon and behave as free Bunch-Davies modes. At late times, when $|k\tau|\ll 1$, the evolution is controlled by $\delta_n$, which depends on the physical mass scale of the tower. For $\delta_n < 9/4$, the order of the Hankel function is real, whereas for $\delta_n > 9/4$ it becomes purely imaginary. We thus identify two cases depending on the mass.

 \subsubsection{Heavy Modes}
\label{sec:heavymodes}

Modes with $\delta_n > 9/4$ correspond to masses above the Hubble scale $nm_t > \tfrac{3}{2}H$, see eq.~\eqref{eq:delta_def}. Their late–time evolution is governed by a Hankel function of purely imaginary order, together with the time–dependent prefactors of the full solution, so they continue to oscillate after horizon exit with a characteristic logarithmic modulation. Moreover, their amplitude is exponentially suppressed, and this suppression becomes even stronger for heavier modes, which have larger imaginary order.

Since these heavy modes remain oscillatory outside the horizon, it is convenient to describe them in terms of Bogolyubov coefficients \cite{Kofman:1997yn}, which parameterize the mixing between positive- and negative-frequency components of the solution. Their explicit form is presented in Appendix~\ref{sec:appendix_modes}, where the large-$\delta_n$ solution for $\xi_n$ is given. We are now interested to compute the two-point correlation function, which tells us how fluctuations of a field at different points in space (or momentum) are correlated. The result in position space for a single heavy mode is
\begin{equation}
\label{eq:heavy_corr_maintext}
\left\langle \chi_n(\tau,\vec{x})\, \chi_n(\tau,\vec{x}) \right\rangle
\simeq \dfrac{1}{a^2}\,\frac{e^{-2 \pi  \sqrt{\delta_n}}\,\delta_n}{6 \pi ^2 \tau ^2}
= \frac{e^{-2 \pi  n/N_H }\,m_1^2 e^{-2\gamma \varphi}\, n^2}{6 \pi ^2} \; .
\end{equation}
The contribution of these modes decays exponentially with their mass, and they will generally give a subdominant contribution compared to the light modes. There are then two possibilities for the heavy modes. If the  mass of the tower sits below the Hubble scale such that the massgap of the heavy modes is less than $H$, the entire tower will have a contribution 
	\begin{equation}
		\sum_{n=3N_H/2+1}^N\dfrac{m_n^2}{2} e^{-2\gamma \varphi}
		\langle\chi_n(\tau, \vec{x})\, \chi_n(\tau, \vec{x})\rangle\simeq\dfrac{4\times10^{-5}}{\pi^2}H^4 N_H,
        \end{equation}
with $N\gg N_H$ some arbitrarily high cutoff \footnote{We can formally take this cutoff to infinity, such that we \textit{integrate in} the entire tower of states. This is in agreement with the regularization procedure introduced in \cite{Blumenhagen:2023tev, Blumenhagen:2024ydy}}. Then, this contribution is parametrically the same as that on the light modes, but numerically suppressed, and can be neglected. If the tower sits above the Hubble scale such that there are no light modes, the contribution to the inflaton dynamics is 
  \begin{equation}
		\sum_{n=1}^N\dfrac{m_n^2}{2} e^{-2\gamma \varphi}
		\langle\chi_n(\tau, \vec{x})\ \chi_n(\tau, \vec{x})\rangle\simeq\dfrac{e^{-2\pi m_t/H}}{12\pi^2} m_t^4<\dfrac{H^4}{12\pi^2}
		\label{eq:twopcorrexph2a} \; ,
	\end{equation}
and will be exponentially suppressed by the mass of the tower. In fact, we can see that in this case the total contribution is less than that of a single light mode. In the following we consider only the contributions from light modes, as we have shown that modes with masses above Hubble do not contribute significantly given our choice of coupling.
    
 Here we should mention that whenever one has a tower of KK scalars from compactification, one also inevitably has a KK tower of massive spin-2 particles, originating from the dimensional reduction of the higher-dimensional graviton. Such a tower will then violate unitarity whenever its mass scale is below the Hubble scale~\cite{Higuchi:1986py}, namely, it will violate the Higuchi bound 
\begin{eqnarray}
    m_{\text{spin-2}}^2\geq 2 H^2 \;  ,
\end{eqnarray} 
where the violation of unitarity is due to a 0-helicity mode acquiring negative kinetic energy. Here we restrict ourselves to a single tower of light scalars, whose degeneracy matches that of a KK tower, but otherwise assume there are no massive gravitons below the Hubble scale. In order to satisfy the Higuchi bound with a light tower of KK scalars we would need to assume that the scale separation between spin-0 and spin-2 modes can effectively arise because only the KK scalars are coupling directly to the inflaton. Through this coupling, their masses can receive additional suppression relative to the spin-2 modes, allowing the EFT to satisfy the Higuchi bound. Additionally, we can also satisfy the Higuchi bound by assuming that the KK scale never decreases below Hubble scale, so that we have only massive modes that do not change the predictions for inflation significantly (see \cite{Scalisi:2019gfv}). In \cite{Lust:2019lmq}, it was argued, for towers of string excitation modes, that UV effects actually prevent the breakdown of the EFT due to a violation of the Higuchi bound. If one tried to embed our setup in string theory one might expect a similar thing, in which one is prevented from lowering the KK scale arbitrarily. In any case we will focus on light modes, which can give a non-negligible contribution and comment on this in the conclusions.

\subsubsection{Light modes}
Modes with $\delta_n \leq 9/4$, corresponding to masses below the Hubble scale $nm_t \leq \tfrac{3}{2}H$, do not oscillate at late times. Notice that the normalization of~\cref{eq:xisol1} is chosen such that modes with very short wavelength, \emph{i.e.} with $k^2 \tau^2 \gg |2-\delta_n| $, match the Bunch-Davies vacuum. Conversely, at late times, for small $(-k \tau)$, the light modes behave as
	\begin{equation}
		\xi_n(\tau, \vec{k})\simeq  \dfrac{i}{\sqrt{2 k } }(-k\tau)^{-1 + \delta_n/3 }\; .
		\label{eq:solxi}
	\end{equation}
	
	Using the definition of $\xi_n$, the two-point correlation function for $\chi_n$ is then given by
	\begin{equation}
		\langle \chi_n(\tau, \vec{k})\text{ } \chi_m(\tau, \vec{k'}) \rangle=\delta_{nm} \, \delta(\vec{k}+\vec{k'}) \dfrac{ |\xi_n(\tau,\vec{k})|^2 }{a^2} \;  ,
		\label{eq:chi_correlation_full}
	\end{equation}
	where $\delta_{nm}$ is a Kronecker delta function. In position space, this equation reads
	\begin{equation}
		\langle \chi_n(\tau, \vec{x})\text{ } \chi_n(\tau, \vec{x}) \rangle \simeq  \dfrac{1}{a^2}\int_{0}^{\frac{\sqrt{(2-\delta_n)}}{|\tau|}} \, |\xi_n(\tau,\vec{k})|^2  \, \dfrac{4\pi k^2 \textrm{d} k}{(2\pi)^3} \simeq \dfrac{3 H^2}{ 8 \pi^2 \delta_n  } \; ,
		\label{eq:chi_correlation}
	\end{equation}
	where, since modes are enhanced when they have a size comparable to the Hubble radius at that time, we have approximated the $k$-integral by substituting~\cref{eq:solxi}. Moreover, we have further approximated the integral using $\delta_n\ll1$. This condition is related to the following constraint that we impose in our computation.

\begin{tcolorbox}[title=Constraint C3: Number of modes]
In our analysis, we impose that the number of light modes is large, namely
\begin{equation}
N_H \gg 1 \; .
\label{eq:expca3}
\end{equation}
Comparing with~\cref{eq:NH}, this requirement is equivalent to
\begin{equation}
m_1 e^{-\gamma \varphi/M_{\rm P}} \ll H \; ,
\label{eq:number_of_modes}
\end{equation}
and, using~\cref{eq:delta_def}, it also implies
\begin{equation}
\delta_n \ll 1 \; ,
\end{equation}
a limit, which approximately reproduces the standard freezing behavior of massless modes. For computational purposes, we then require that at least ten modes contribute to the inflationary dynamics, ensuring that the large-$N_H$ limit can be taken reliably. Unlike the previous constraints, which follow from physical considerations, this assumption is not essential for the consistency of the framework. It is simply a practical choice that simplifies our numerical analysis and does not stem from high-energy physics arguments. Consequently, one could relax this assumption and extend the results of this work to scenarios in which only a few modes in the tower are active during inflation.
\end{tcolorbox}

\begin{figure}[t]
		\centering\includegraphics[width=0.6\textwidth]{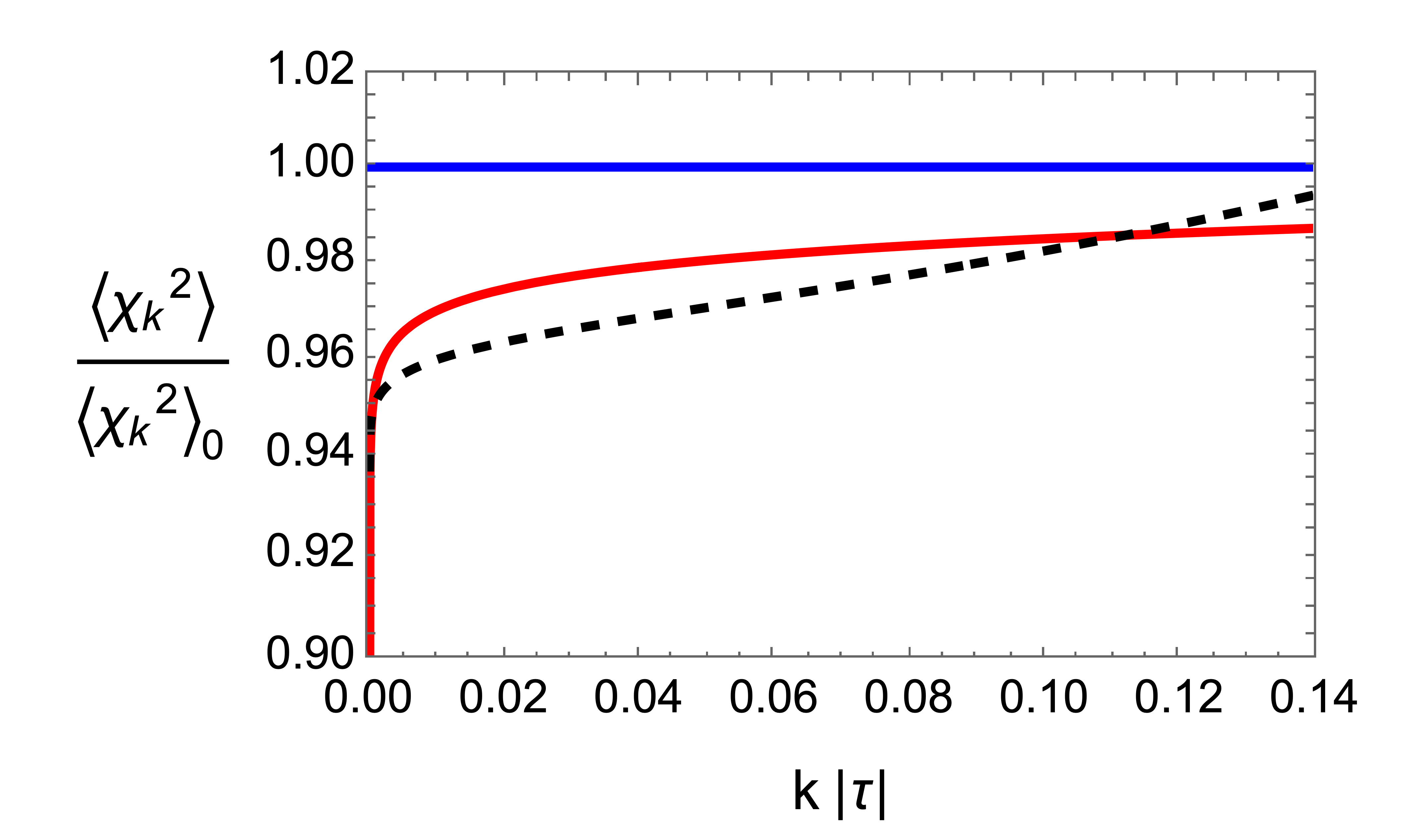}
		\caption{Two-point correlation function for the $\chi$ fields, compared to the expected value of vacuum modes (solid blue line). The solid red line denotes the approximate solution of~\cref{eq:xi_eq_2} (divided by the scale factor) for $k|\tau|\ll\sqrt{2-\delta_n}$, while the dashed black line is the numerical solution. Here we have fixed $\delta_n=0.01$, $\gamma=0.25$, $|\dot{\vp}|/H=1/\sqrt{120}$, corresponding to the single-field solution of chaotic inflation with a linear potential.}
		\label{fig:twopointcorr}
	\end{figure}

   In~\Cref{fig:twopointcorr}, we compare the result of~\cref{eq:chi_correlation_full}, obtained using the approximate solution of~\cref{eq:xi_eq_2} for $\delta_n \ll 1$ (red line), with the standard freezing solution (blue line, corresponding to $\delta_n = 0$). We also show the numerical solution of~\cref{eq:xi_eq_2} (black dotted line) and verify that both the approximate and the numerical solutions depart from the freezing behaviour. As we can see, for $(-k\tau)\rightarrow 0$, corresponding to late times, the solution decays instead of freezing.

    \subsection{Backreaction on background dynamics}
    \label{sec:backreaction}
 	
    We proceed by computing the backreaction due to particle production on the inflaton’s dynamics. For this purpose, we  sum over the light modes
	\begin{equation}
		\sum_{n=1}^{3 N_H / 2}\dfrac{m_n^2}{2} e^{-2\gamma \varphi }
		\langle \chi_n(\tau, \vec{x})\text{ } \chi_n(\tau, \vec{x}) \rangle\; \simeq \dfrac{3 H^2}{16 \pi^2} \sum_{n=1}^{3 N_H / 2} \, e^{-2\gamma \varphi }  \, \dfrac{m_n^2  }{ \delta_n} 
		\simeq \dfrac{9 N_H H^4 }{32 \pi ^2}  \;  ,
		\label{eq:twopcorrexp0} 
	\end{equation}
	where we have used~\cref{eq:delta_def} and the sum goes up to $3 N_H / 2$ because of the definition of light modes given in the previous section. Notice that the factor of $\delta_n$ in~\cref{eq:chi_correlation} precisely cancels the exponential prefactor and the explicit dependence on $\gamma$. Thus, the two cases of an exponentially growing or decreasing mass of the tower lead to the same result. At this point, it is worth stressing again that, regardless of whether the mass of the tower is decreasing or increasing, we always assume (and in fact our computation requires) that there is always a large amount of light fields with masses below Hubble (see Constraint C3). However, as we show in the following, the sign of the coupling appears explicitly in some of the perturbative corrections to single-field inflation quantities.
     
     It is convenient to use~\cref{eq:NH_Lambda_eq} to recast~\cref{eq:twopcorrexp0} as
		\begin{equation}
		\sum_n^{N_H}\dfrac{m_n^2}{2} e^{-2\gamma \varphi }
		\langle \chi_n(\tau, \vec{x})\text{ } \chi_n(\tau, \vec{x}) \rangle\;
		\simeq \dfrac{9 H^2 }{32 \pi ^2}\left(\dfrac{H}{\Lambda_{\text{sp}}}\right)^3  \;  .
		\label{eq:twopcorrexp} 
	\end{equation}
    Here we can see that as long as the EFT description remains valid, the contribution to the energy density of the tower will be suppressed with respect to the Hubble scale. Following a similar procedure, we can show that the term depending on $\langle \dot{\chi_n}^2 \rangle$ appearing in~\cref{eq:H_backreactions} is suppressed by a factor $\delta_n$ (\emph{i.e.}, $1/N_H^2$) with respect to the term in~\cref{eq:twopcorrexp0}. Thus, we proceed by neglecting this contribution.

	To estimate the correction to Friedman equation due to the $\chi_n$ fields, we assume slow-roll and approximate~\cref{eq:H_backreactions} as
	\begin{equation}
		\label{eq:H_backreactions_two}
		H^2\simeq\dfrac{V(\vp)}{3 }\left(1+ \sum_n \dfrac{m_n^2}{2} e^{-2\gamma \varphi } \dfrac{\langle\chi_n\chi_n\rangle}{V(\vp)}\right) \simeq \dfrac{V(\vp)}{3 }\left[1+ \dfrac{9}{32 \pi ^2}  \dfrac{H^2 }{V(\vp)}\left(\dfrac{H}{\Lambda_{\text{sp}}}\right)^3\right] \; .
	\end{equation}
	If the second term in the round brackets is much bigger than one, this equation reduces to $N_H H^2 \simeq 16 \pi^2 $, which is inconsistent with Constraint C1 in~\cref{eq:expca22}. Thus, background dynamics dominated by particle production would immediately lead to a breakdown of the EFT. On the other hand, in the limit $3 H^2 \simeq V(\varphi)$ the second term in the bracket reads
	\begin{equation}
		\dfrac{9}{32 \pi ^2}  \dfrac{H^2 }{V(\vp)}\left(\dfrac{H}{\Lambda_{\text{sp}}}\right)^3 \simeq 
		\dfrac{3}{32 \pi ^2}  \left(\dfrac{H}{\Lambda_{\text{sp}}}\right)^3 < 0.0095   \;  ,
	\end{equation}
	where in the inequality we have used the Constraint C1, namely that $H<\Lambda_{\rm sp}$. This shows that, indeed, the contribution of the $\chi_n$ fields to the Friedman equation is negligible, which justifies approximating the Friedman equation as $3 H^2\simeq V(\vp) $.

	Finally, assuming slow-roll,~\cref{eq:eom_varphi_backreactions} reduces to
    \begin{equation}
		\label{eq:eom_varphi_final}
		3 H \dot\varphi + V_{, \varphi}  -  \sum_n^{N_H} \gamma \, m_n^2 \,  e^{-2\gamma \varphi } \langle\chi_n \chi_n \rangle  \simeq 3 H \dot\varphi + V_{, \varphi} - \frac{9\gamma}{16 \pi^2}H^2 \left(\dfrac{H}{\Lambda_{\text{sp}}}\right)^3 \simeq 0 \; .
	\end{equation}
	While the first term corresponds to the usual Hubble friction, the next two terms are the force exerted by the potential and the $\chi_n$ fields, respectively. 
 The field velocity can be written as
 \begin{equation}
     \label{fieldvel}
     \dot\vp=-\dfrac{V_{,\vp}}{3 H}\left[1-\dfrac{3 \gamma}{16\pi^2}\left(\dfrac{H}{\Lambda_{\text{sp}}}\right)^3 \left(  \frac{ \textrm{d} \ln V  }{ \textrm{d} \vp } \right)^{-1} \right] \; .
 \end{equation}
For later convenience, we introduce a fourth constraint.

\begin{tcolorbox}[title=Constraint C4: Field velocity]
In our analysis, we impose that the tower contribution to the field velocity is negligible, namely
 	\begin{equation}
		 \frac{3\gamma}{16 \pi^2} \left(\dfrac{H}{\Lambda_{\text{sp}}}\right)^3 \left(  \frac{ \textrm{d} \ln V  }{ \textrm{d} \vp } \right)^{-1} \ll1  \;  .
		\label{eq:field_velocity}
	\end{equation}
This constraint ensures that the last part of the inflationary trajectory (\emph{i.e.}, from CMB scales to much smaller scales) is only marginally affected by the presence of the $\chi_n$ fields. In the following section, we will explicitly check the validity of this approximation for the models of our interest and find that, in all our examples, this condition is essentially guaranteed by the other three constraints.

\end{tcolorbox}

We conclude this section by introducing the modified expressions for the slow-roll parameters in~\cref{eq:epsilon,eq:eta}
\begin{align}
   \label{eq:ve_approximated}
		\ve  \simeq & \, \frac{1}{2} \left( \frac{\textrm{d} \ln V}{\textrm{d} \vp} \right)^2  - \dfrac{3\gamma  }{32 \pi ^2} \left(\dfrac{H}{\Lambda_{\text{sp}}}\right)^3  \left(  \frac{ \textrm{d} \ln V  }{ \textrm{d} \vp } \right)     \; , \\
    \label{eq:eta_approximated}
		\eta  \simeq & \, \frac{ V_{,\varphi \varphi}}{V} - \frac{1}{2} \left(  \frac{ \textrm{d} \ln V  }{ \textrm{d} \vp } \right)^2 - \dfrac{9\gamma  }{32 \pi ^2}  \left(\dfrac{H}{\Lambda_{\text{sp}}}\right)^3 \left(  \frac{ \textrm{d} \ln V  }{ \textrm{d} \vp } \right)    \; . 
\end{align}
For later convenience, we denote with $\ve_0$ and $\eta_0$ the standard slow-parameters, without the tower contribution, and with $\Delta \ve$, $\Delta \eta$ the corrections induced by the tower of scalars, \emph{i.e.},
\begin{equation}
\begin{aligned}
\label{eq:slowrollexpanded}
    \ve_0 = \frac{1}{2} \left( \frac{\textrm{d} \ln V}{\textrm{d} \vp} \right)^2 \;,  \qquad\,\,\,\,  \eta_0 =  \, \frac{ V_{,\varphi \varphi}}{V} - \frac{1}{2} \left(  \frac{ \textrm{d} \ln V  }{ \textrm{d} \vp } \right)^2 \; ,\qquad  \\
    \Delta \ve = -\dfrac{3\gamma  }{32 \pi ^2} \left(\dfrac{H}{\Lambda_{\text{sp}}}\right)^3  \left(  \frac{ \textrm{d} \ln V  }{ \textrm{d} \vp } \right) \;,  \qquad  \Delta \eta = -\dfrac{3\gamma  }{8 \pi ^2} \left(\dfrac{H}{\Lambda_{\text{sp}}}\right)^3  \left(  \frac{ \textrm{d} \ln V  }{ \textrm{d} \vp } \right) \; . 
\end{aligned}
\end{equation}
Notice that these corrections are of the same order of the field velocity constraint in~\cref{eq:field_velocity} with respect to $\ve_0$ and $\eta_0$.

\subsection{Perturbations and observable quantities}
	\label{sec:perturbations}
	The presence of the $\chi_n$ fields not only affects the background evolutions, but it also modifies the scalar and tensor power spectra, which, in the following, we will express as
	\begin{equation}
		P_{\zeta} = P^{\varphi}_{\zeta} + P^{\rm \chi}_{\zeta} \; ,  \qquad P_{T} = P^{\varphi}_{T} + P^{\chi}_{T} \; ,
	\end{equation}
	where $P^{\varphi}_{\zeta}$ and $P^{\varphi}_{T}$ are the single-field contributions to the scalar and tensor power spectra reported in~\cref{eq:power_spectra_vacuum}, and $P^{\chi}_{\zeta}$, $P^{\chi}_{T}$ are the contributions sourced by the $\chi_n$ fields. Let us start by focusing on the scalar perturbations. The e.o.m for $\delta \vp$, in momentum space and in terms of $\psi \equiv a \delta\vp $, reads
	\begin{equation}
		\begin{aligned}
			\psi'' + \left[k^2 -\dfrac{a''}{a} +a^2\dfrac{\partial^2 V(\varphi)}{\partial\varphi^2} +2 \sum_n \dfrac{\gamma^2 a^2 m_n^2 }{M_{\rm P}}  e^{ -2\gamma \varphi }  \langle\chi_n\chi_n\rangle \right] \psi = \sum_n  \gamma a^3 m_n^2   e^{ -2\gamma \varphi } \delta (\langle\chi_n\chi_n\rangle)  \;  ,
		\end{aligned}
	\end{equation}
	where $\delta (\langle\chi_n\chi_n\rangle) = \chi_n\chi_n - \langle\chi_n\chi_n\rangle$ . Expanding to the lowest order in slow-roll parameters and using~\cref{eq:twopcorrexp}, this equation can be written as
	\begin{equation}
		\psi''+\left[ k^2 -\dfrac{2 - \beta^2 }{\tau^2} \right]\psi = \sum_n \dfrac{\delta_n \gamma  }{ H (-\tau)^{3}}  \delta (\langle\chi_n\chi_n\rangle)  \; ,
		\label{eq:sourcedeq}
	\end{equation}
	where we have used~\cref{eq:delta_def}, and to ease the notation we have introduced
	\begin{equation}
		\beta^2 \equiv \dfrac{9 \gamma^2 }{8 \pi ^2} \left(\dfrac{H}{\Lambda_{\text{sp}}}\right)^3  \; . 
	\end{equation} 
    Notice that the Constraint C1 in~\cref{eq:expca2} implies $\beta \ll 1$. The full solution $\psi$ is the sum of $\psi_h$, the solution of the homogeneous equation, and $\psi_s$, the solution of the sourced equation. The homogeneous solution is given by 
	\begin{equation}
		\psi_h(\tau, \vec{k})=\frac{\sqrt{\pi }}{2} \sqrt{-\tau} \,H_{\Delta}^{(1)}(-k \tau) 
	\end{equation}
	where $\Delta \equiv \frac{1}{2} \sqrt{9-4\beta^2}$ and $H_{\Delta}^{(1)}$ is the Hankel function. The single-field contribution can then be expressed as
	\begin{equation}
		P^{\varphi}_{\zeta} =\dfrac{k^3}{ 2\pi^2} \, \dfrac{H^2}{\dot\vp^2 }\left|\delta\vp_h \right|^2 \simeq \dfrac{1}{ 4\pi^2} \, \dfrac{H^4 }{\dot\vp^2 }(-k\tau)^{ 2\beta^2 /3 } \simeq \dfrac{1}{ 4\pi^2} \, \dfrac{H^4 }{\dot\vp^2 } \; ,  
	\end{equation}
	where we have applied the limit $\beta\ll 1$ in the last equality.  This coincides with the usual result for single-field inflation with a standard kinetic term. On the other hand, the term $\psi_s$ sourced by the $\chi_n$ fields can be computed using the Green function for~\cref{eq:sourcedeq} and convolving with the source term (see Appendix~\ref{sec:corrfuncs} for details). 
	Using~\cref{eq:sourced_spectrum_final} the total scalar spectrum reads
	\begin{equation}
    \label{eq:scalar_spectrum}
		P_{\zeta}  
		 = \frac{H^4}{4 \pi^2 \dot{\vp}^2}  \left[ 1 +  \frac{1}{8 \pi^2}  \dfrac{ \gamma^2   }{ 27    }  \left(\dfrac{H}{\Lambda_{\text{sp}}}\right)^3 \right] \; ,
	\end{equation}
 where the second term, sourced by the tower, is smaller than unity, once we assume that $\gamma=\mathcal{O}(1)$ as suggested by eq.~\eqref{eq:sdconst}. This implies that the contribution to the scalar power spectrum sourced by the $\chi_n$ fields is typically negligible. The scalar spectral index defined in~\cref{eq:nsr00} then reads
\begin{equation}
 \label{eq:ns_approximated}
	n_s \simeq 1 + 2 (\eta_0+\Delta\eta)  -4  (\ve_0+\Delta\ve ) -  \frac{\gamma^2}{72 \pi^2}   \left(\dfrac{H}{\Lambda_{\text{sp}}}\right)^3  \ve_0 \; ,
	\end{equation}
    where we separate the contributions to the slow-roll parameters as in~\cref{eq:slowrollexpanded}.
Substituting the modified expressions of $\ve$ and $\eta$ we get 
	\begin{equation}
		n_s \simeq 1 +\eta_0 - 2 \epsilon_0
        -   \frac{3 \gamma}{16 \pi^2  } \sqrt{2\epsilon_0 }  \left(\dfrac{H}{\Lambda_{\text{sp}}}\right)^3    
        \; ,
		\label{eq:ns_final}
	\end{equation}
implying that the correction induced by the $\chi_n$ fields is typically suppressed with respect to the single-field term. 
This means that, in the regime in which the Hubble and QG scales remain separated, the contributions of the tower remain negligible. When the scales become comparable $H/\Lambda_{\text{sp}}\geq \mathcal{O}(10^{-1})$, one may have sizeable contributions for $\gamma\geq \mathcal{O}(1)$. Notice that the limit $P^{\chi}_{\zeta} / P^{\varphi}_{\zeta} \rightarrow 0$ corresponds to $m_{\text{t}}\rightarrow \infty$, implying that the tower decouples completely from the inflaton.

	Similarly, we can show that the equation of motion for tensor perturbations reads
	\begin{equation}
		\gamma_{ij}^{\prime \prime} + \left( k^2 - \frac{2}{\tau^2} \right) \gamma_{ij} = a(\tau) \Pi_{ij}{}^{ab}( \vec{k} )\,T_{ab}(\vec{k},\,\tau') \; , 
		\label{eq:eom_tensors}
	\end{equation}
	where $\gamma_{ij} \equiv a h_{ij} / 2$, $\Pi_{ij}{}^{ab}(\hat{k})$ is the traceless transverse projector
 \begin{equation}
     \Pi_{ij}{}^{ab}(\hat{k}) \equiv \Pi_{i}^{a} (\hat{k}) \, \Pi_{j}^{b}(\hat{k})  - \frac{1}{2} \Pi_{ij} (\hat{k})\, \Pi_{ab}(\hat{k}) \;, \qquad \Pi_{ij}(\hat{k}) \equiv \delta_{ij} - \hat{k}_i \hat{k}_j \;, 
 \end{equation} 
 and $T_{ab}$ is the stress energy tensor for the $\chi_n$ fields. The tensor power spectrum, including the sourced term, is given by (for details, see Appendix~\ref{sec:corrfuncs}, and in particular,~\cref{eq:tensor_spectrum_app})
	\begin{equation}
		\label{eq:tensor_spectrum}
		P_{T} = \dfrac{2\,H^2}{\pi^2}\,\left[1+\frac{2.3}{10^{3}} \left(\dfrac{H}{\Lambda_{\text{sp}}}\right)^3\right] ,
	\end{equation}
 where the first and second terms are coming from the inflaton and the tower fields, respectively.
 From~\cref{eq:scalar_spectrum,eq:tensor_spectrum} we find the tensor-to-scalar ratio
 \begin{equation}
 \label{eq:r_approximated}
     r\simeq 16\ve_0\left[1-\frac{1}{\sqrt{2\ve_0}}\frac{3\gamma}{16\pi^2}\left(\dfrac{H}{\Lambda_{\text{sp}}}\right)^3\right] \; .
 \end{equation}
As for the scalar power spectrum, $H/\Lambda_{\text{sp}} \ll 1 $ implies that the $\chi_n$-sourced contribution is negligible and the expression of the tensor-to-scalar ratio is mostly unchanged compared to~\cref{eq:nsr00}.

	We conclude this section by computing the non-linearity parameter in~\cref{eq:FNL_def}. For this purpose, we use~\cref{eq:three_point_final} to express the bispectrum in the equilateral limit
	\begin{equation}
		B^{\rm equil}_\zeta (k_1,k_2,k_3)=-\dfrac{H^6}{\dot{\vp_0}^3}   \left( \frac{2 \gamma H }{9  } \right)^3 \dfrac{3}{64\sqrt{2}\pi^{7/2}k^6} \frac{N_H}{5}  \;  ,
	\end{equation}
	which implies that the sourced contribution to $f^{\rm equil}_{\rm NL}$ is
	\begin{equation}
		f^{\rm equil}_{\rm NL}\simeq  \dfrac{\gamma^3 }{2916 \pi^2} \,   \sqrt{ 2 \ve_0 } \,  \left(\dfrac{H}{\Lambda_{\text{sp}}}\right)^3 \;  .
	\end{equation}
    The non-gaussianity parameter vanishes as one decouples the tower from the inflaton, either by taking $\gamma\to 0$ or $\Lambda_{\text{sp}}\to\infty$, which is the expected result for single-field inflation.
	For $\gamma\simeq H/\Lambda_{\text{sp}}\simeq \mathcal{O}(1)$, and using typical values for $\ve_0$ we can estimate the sourced contribution to $f^{\rm equil}_{\rm NL}$ to be
    \begin{equation}
       f^{\rm equil}_{\rm NL}\lesssim 10^{-6} \;  ,
    \end{equation}
	which is compatible with the experimental bounds~\cite{Planck:2019kim}.

	\section{Implications for inflationary phenomenology}
    \label{sec:pheno}

    In this section, we consider several representative classes of inflationary models to assess the impact of the tower on the CMB observables. In particular, we consider power law inflation~\cite{Lucchin:1984yf}, chaotic inflation~\cite{Linde:1983gd}, $\alpha$-attractor models~\cite{Kallosh:2013yoa,Galante:2014ifa,Roest:2015qya}, including Starobinsky inflation~\cite{Starobinsky:1980te}, and inverse hilltop inflation~\cite{Broy:2015qna}. Without loss of generality, we assume a positive inflaton field ($\vp > 0$) to roll down a potential with $\dot\vp \lt 0$. 

	For a wide class of inflationary models, we can parameterize the slow-roll parameters in terms of the number of e-folds as~\cite{Mukhanov:2013tua, Roest:2013fha, Garcia-Bellido:2014wfa, Binetruy:2014zya}
	\begin{equation}
		\label{eq:ve_universal}
		\ve \simeq \dfrac{\beta_\alpha}{N_{e}^{\alpha}} \;,\qquad \qquad \eta \simeq -\dfrac{\alpha}{2 N_{e}}  + \dfrac{\beta_\alpha}{N_{e}^{\alpha}} \; ,
	\end{equation}
	where $\alpha$ and $\beta$ are positive and order-one constants. In this parameterization, $\alpha$ specifies the model, and $\beta$ sets some of the model parameters. In particular, we have Power Law Inflation models for $\alpha=0$,\footnote{One could in principle also consider potentials of the form $V(\phi)=V_0 e^{\lambda \phi^{1/p}}$, with $p>1$. These lead to $\alpha=1-\frac{1}{2 p-1}$, interpolating between Power-law and Chaotic inflation.} Chaotic Inflation models for $\alpha=1$, Exponentially flat potentials for $\alpha=2$, Inverse Hilltop for $1<\alpha<2$, and Hilltop for $\alpha>2$. Notice that all large field models, \emph{i.e.}, models with field excursion $\Delta \varphi \simeq M_{\rm P} $, have $\alpha \leq 2$. Substituting~\cref{eq:ve_universal} into~\cref{eq:ns_final} and~\cref{eq:r_approximated} we get \begin{eqnarray}\label{eq:ns_final_parameterization}
		n_s & \simeq & 1- \frac{\alpha}{N_e}  - 2 \frac{\beta_\alpha}{N_e^\alpha} 
        -   \frac{3 \gamma}{16 \pi^2  } \left(\dfrac{H}{\Lambda_{\text{sp}}}\right)^3 \frac{ \sqrt{2\beta_{\alpha}} }{  N_e^{\alpha/2 } }     
        \; , \\
        \label{eq:r_final_parameterization}
		r & \simeq & \dfrac{16\beta_{\alpha}}{N_e^{\alpha}}\left[1-\dfrac{N_{e}^{\alpha/2}}{\sqrt{2 \beta_{\alpha}}}\frac{3\gamma}{16\pi^2}\left(\dfrac{H}{\Lambda_{\text{sp}}}\right)^3\right] \; .
	\end{eqnarray}
    In Appendix~\ref{sec:appendix_models}, we report the potentials, which are also summarized in~\Cref{tab:modelstable}, and the modified expressions for $n_s$ and $r$ for all the models considered in this work.

 	\renewcommand{\arraystretch}{2.2}
\begin{table}[htb]
    \centering
    \begin{tabular}{|c|c|c|c|}
    \hline \textbf{Model class}& \textbf{Potential $V(\vp)$}  & $\alpha$
      & $\beta_\alpha$\\ \hline
      Power Law   &$V_0\, e^{\lambda\vp}$ &0  &$\dfrac{\lambda^2}{2}$\\\hline
   Chaotic &  $V_0\, \vp^{p}$  & 1 &$\dfrac{p}{4}$\\\hline
         Inverse Hilltop&$V_0\left(1-\frac{2(\frac{\lambda}{q})^{-q}}{q-2}\vp^{2-q}\right)^2$ & $2-\dfrac{2}{q}$ & $\dfrac{(4q)^{2/q}}{2\lambda ^2}$\\\hline
     Starobinsky-like&  $V_0\left(1-e^{-\lambda\phi}\right)^2$ &2  &$\dfrac{1}{2 \lambda^2} $\\\hline
    \end{tabular}
    \caption{Scalar potentials considered in this work for $\alpha$ in the range $0\leq\alpha\leq 2$. All parameters appearing in the table are positive numbers.} 
    \label{tab:modelstable}
\end{table}

    Let us proceed by estimating the typical size of the term appearing in the field velocity constraint C4 in~\cref{eq:field_velocity}, which controls the backreaction of the tower on the inflationary dynamics
		\begin{equation}
		\frac{3\gamma}{16 \pi^2} \left(\dfrac{H}{\Lambda_{\text{sp}}}\right)^3  \left(  \frac{ \textrm{d} \ln V  }{ \textrm{d} \vp } \right)^{-1} \simeq \dfrac{ 3\gamma }{16 \pi ^2}  \left(\dfrac{H}{\Lambda_{\text{sp}}}\right)^3  \frac{1}{ \sqrt{2 \ve_0} } \simeq \frac{\gamma}{\sqrt{\beta_\alpha}} \left(\dfrac{H}{\Lambda_{\text{sp}}}\right)^3\left(\frac{N_e^{\alpha/2}}{75 }  \right) \;,		\label{eq:friction_term_comparison}
	\end{equation}
	where we have used~\cref{eq:ve_universal} to express the first slow-roll parameter in terms of the number of e-folds. For $\gamma, \beta_\alpha \sim \mathcal{O}(1) $ and $\alpha \leq 2$, it is clear that the  constraint C1 in~\cref{eq:expca2} implies that the correction induced by the tower is negligible in the last $\sim 60$ e-folds of inflation. 

Our computational scheme will be valid in the region of the parameter space determined by constraints C1-C4~\cref{eq:expca1,eq:expca2,eq:expca3,eq:field_velocity}. In~\Cref{tab:constraints}, we summarize the explicit expressions for these constraints after substituting the parameterization in~\cref{eq:ve_universal}.

	\renewcommand{\arraystretch}{2.8}
\begin{table}[htb]
    \centering
    \renewcommand{\arraystretch}{2.0} 
    \begin{tabular}{|c|c|}
        \hline
        \textbf{Constraint} & \textbf{Condition} \\
        \hline
        C1 & $\displaystyle \frac{H}{\Lambda_{\text{sp}}} < 1$  \\
        \hline
        C2 & $\displaystyle |\gamma| \sqrt{2\beta_\alpha}\, N_e^{-\alpha/2} < 1$ \\
        \hline
        C3 & $\displaystyle \left( \frac{H}{\Lambda_{\text{sp}}} \right) \left( \frac{3}{V(\varphi)} \right)^{1/3} > 1$ \\
        \hline
        C4 & $\displaystyle \frac{3|\gamma|}{16\pi^2} \left( \frac{H}{\Lambda_{\text{sp}}} \right)^3 \frac{N_e^{\alpha/2}}{\sqrt{2\beta_\alpha}} < 1$ \\
        \hline
    \end{tabular}
    \caption{Constraints C1–C4~\cref{eq:expca1,eq:expca2,eq:expca3,eq:field_velocity} used in our computational scheme, expressed using the $\alpha$-parameterization of~\cref{eq:ve_universal}.}
    \label{tab:constraints}
\end{table}

\begin{figure}[t]
    \centering
    \begin{tabular}{ccc}
    \includegraphics[width=0.31\textwidth]{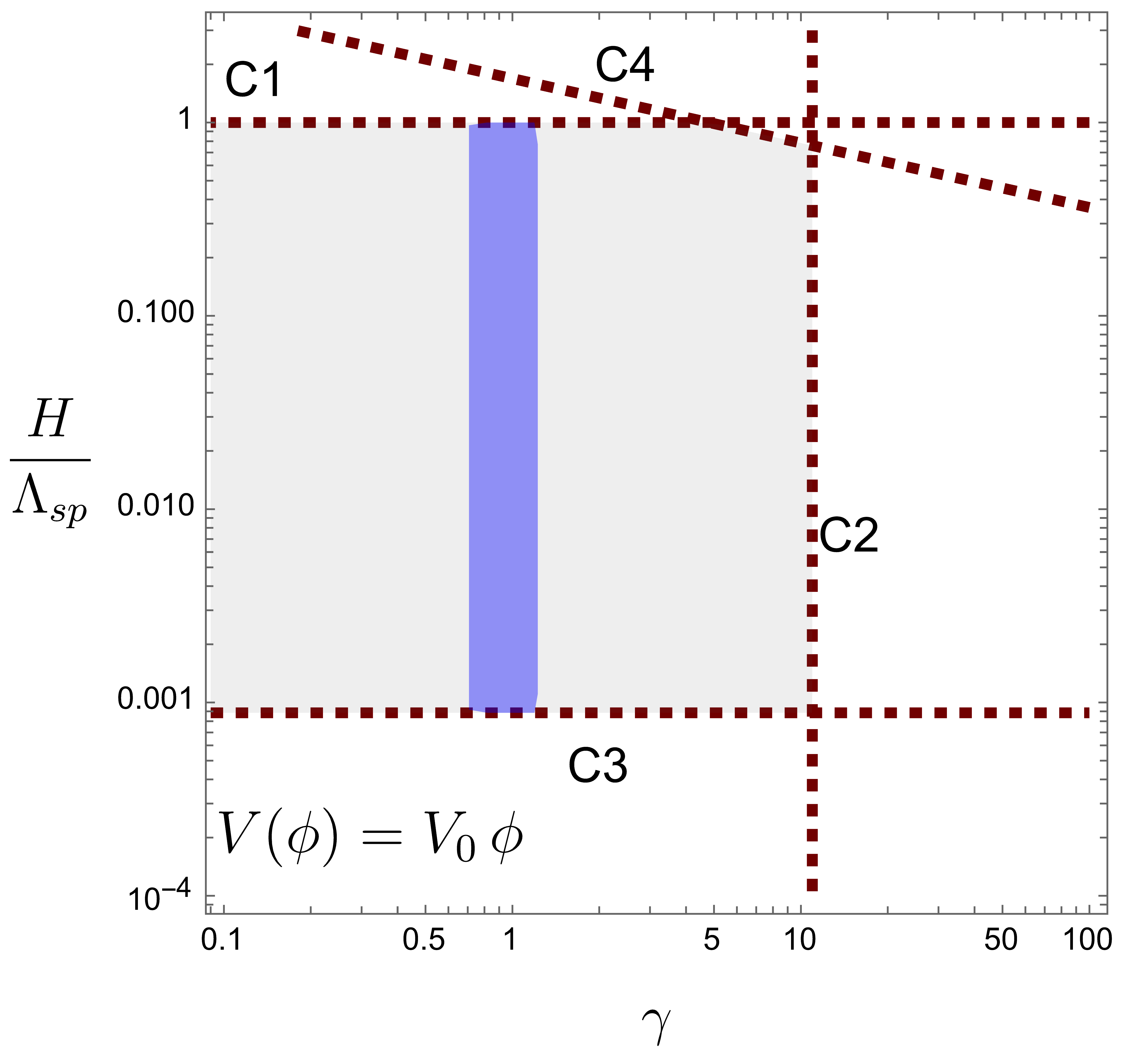} & \includegraphics[width=0.31\textwidth]{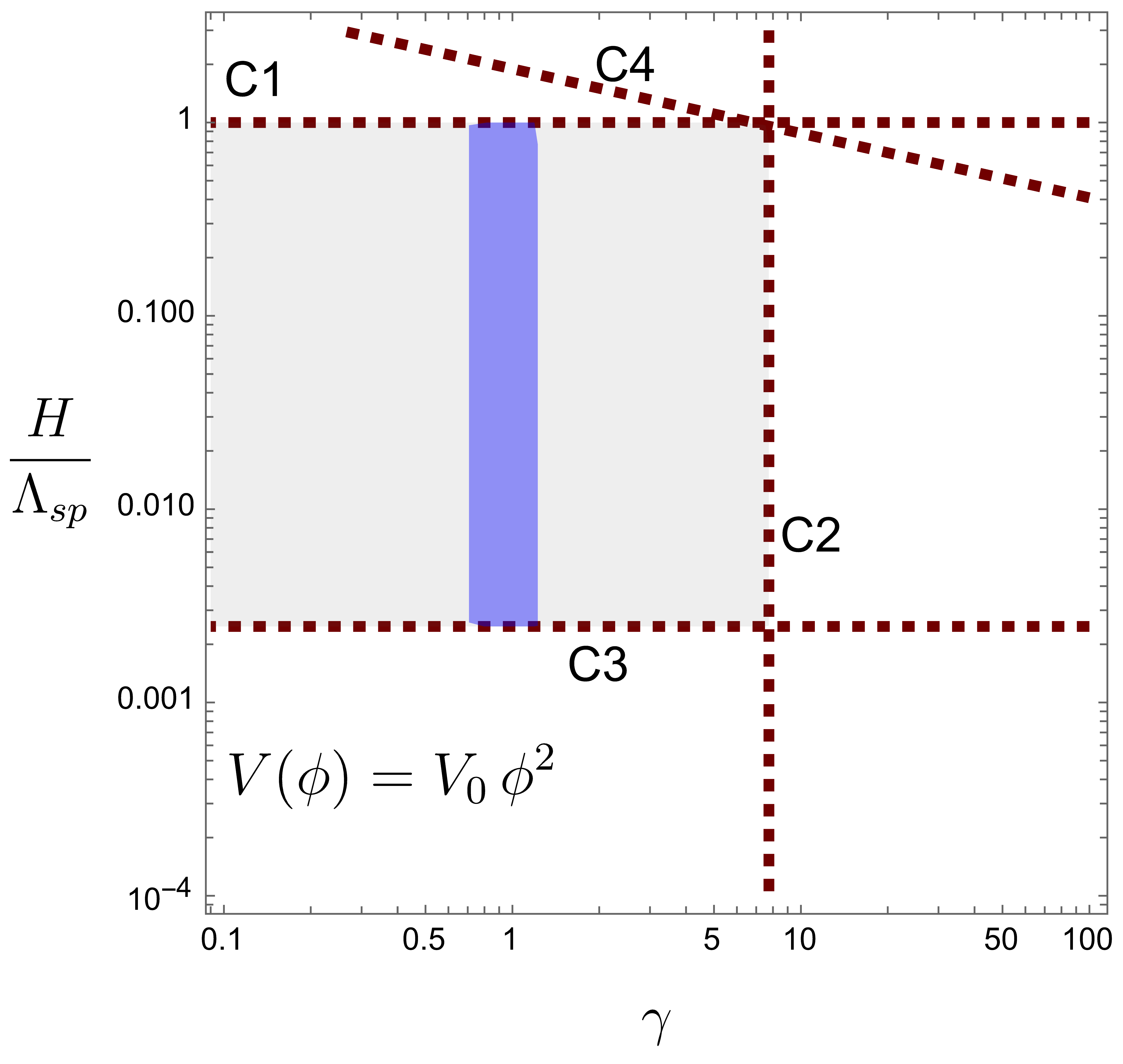} & \includegraphics[width=0.31\textwidth]{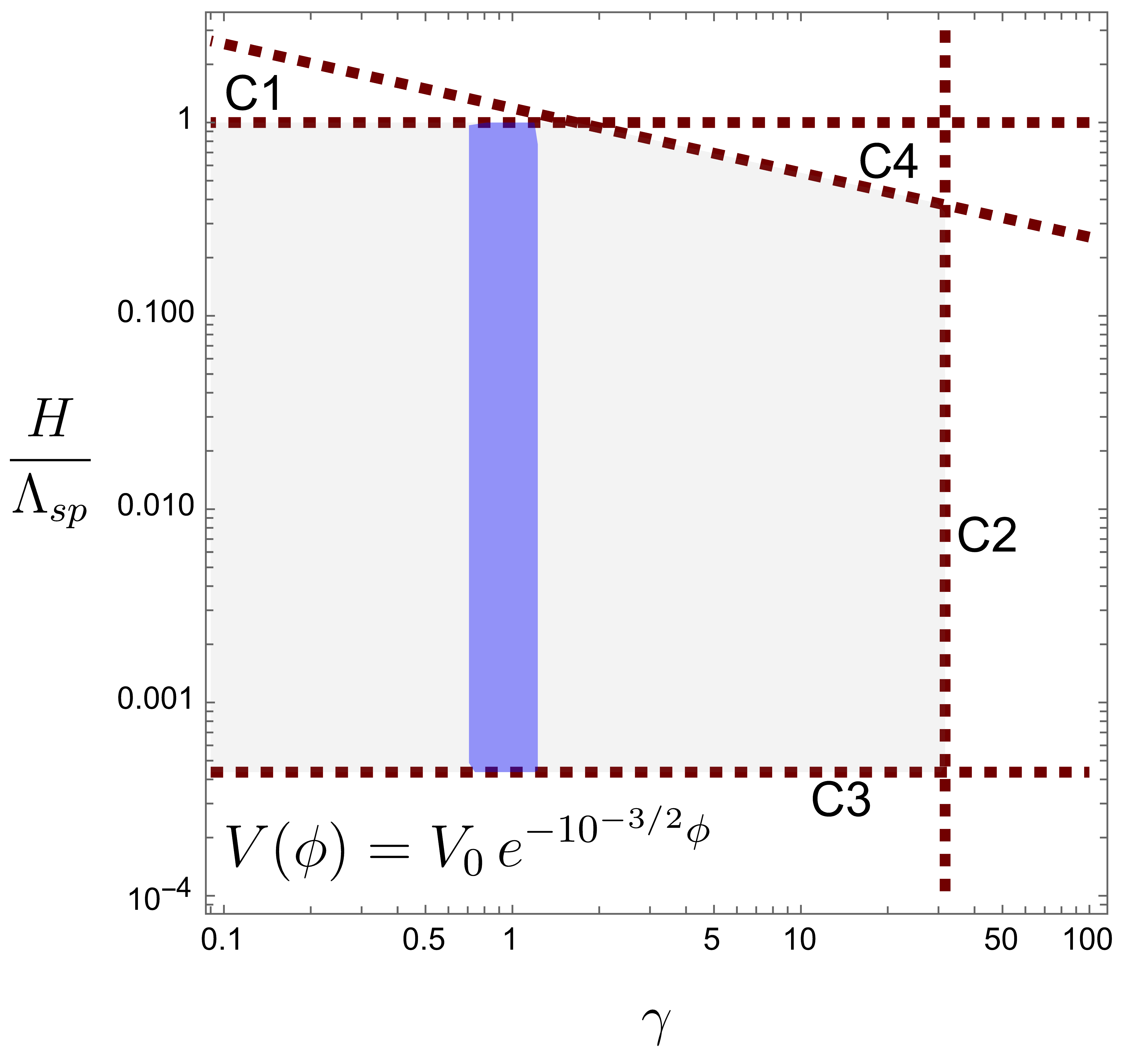}\\ \includegraphics[width=0.31\textwidth]{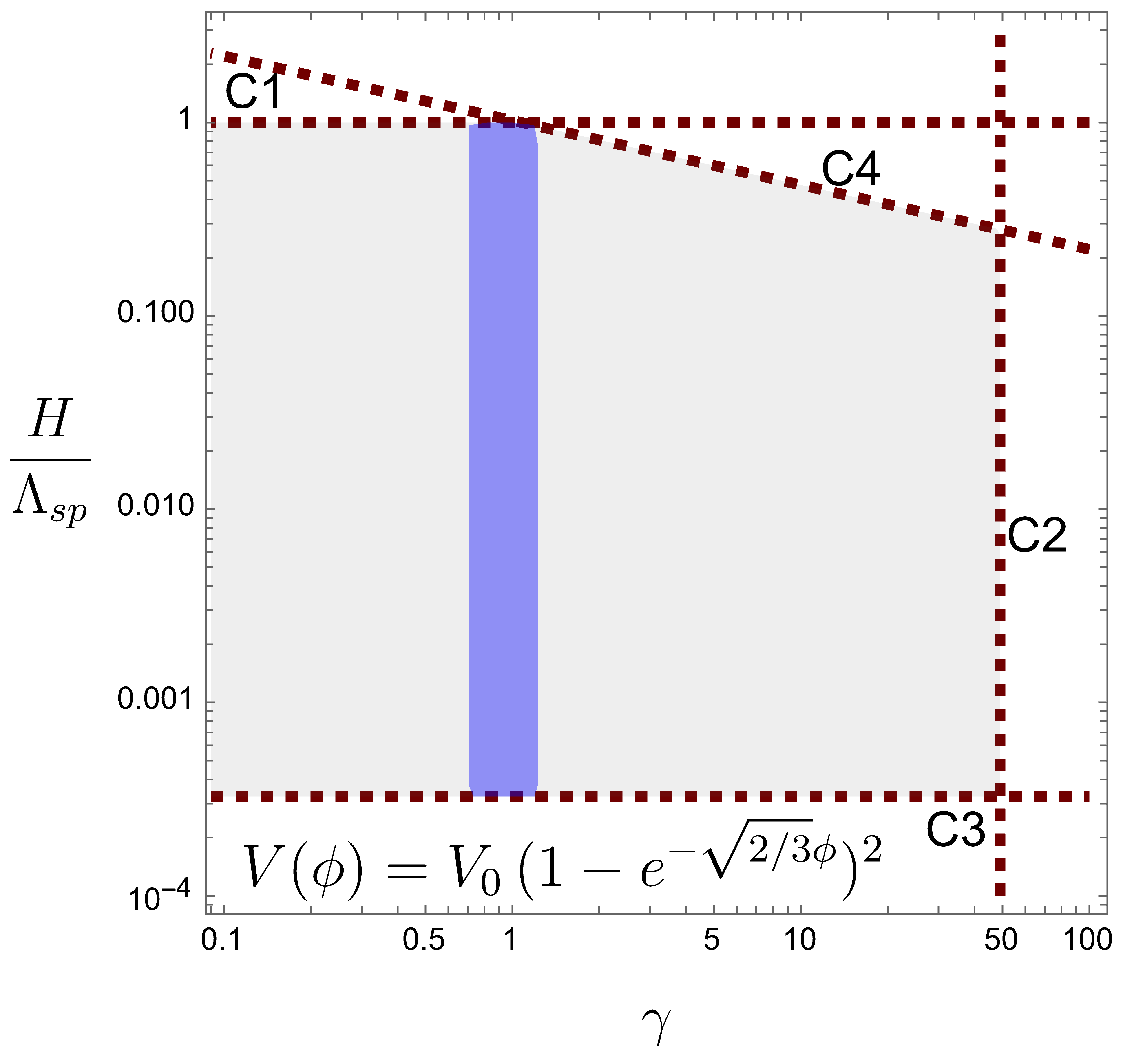} & \includegraphics[width=0.31\textwidth]{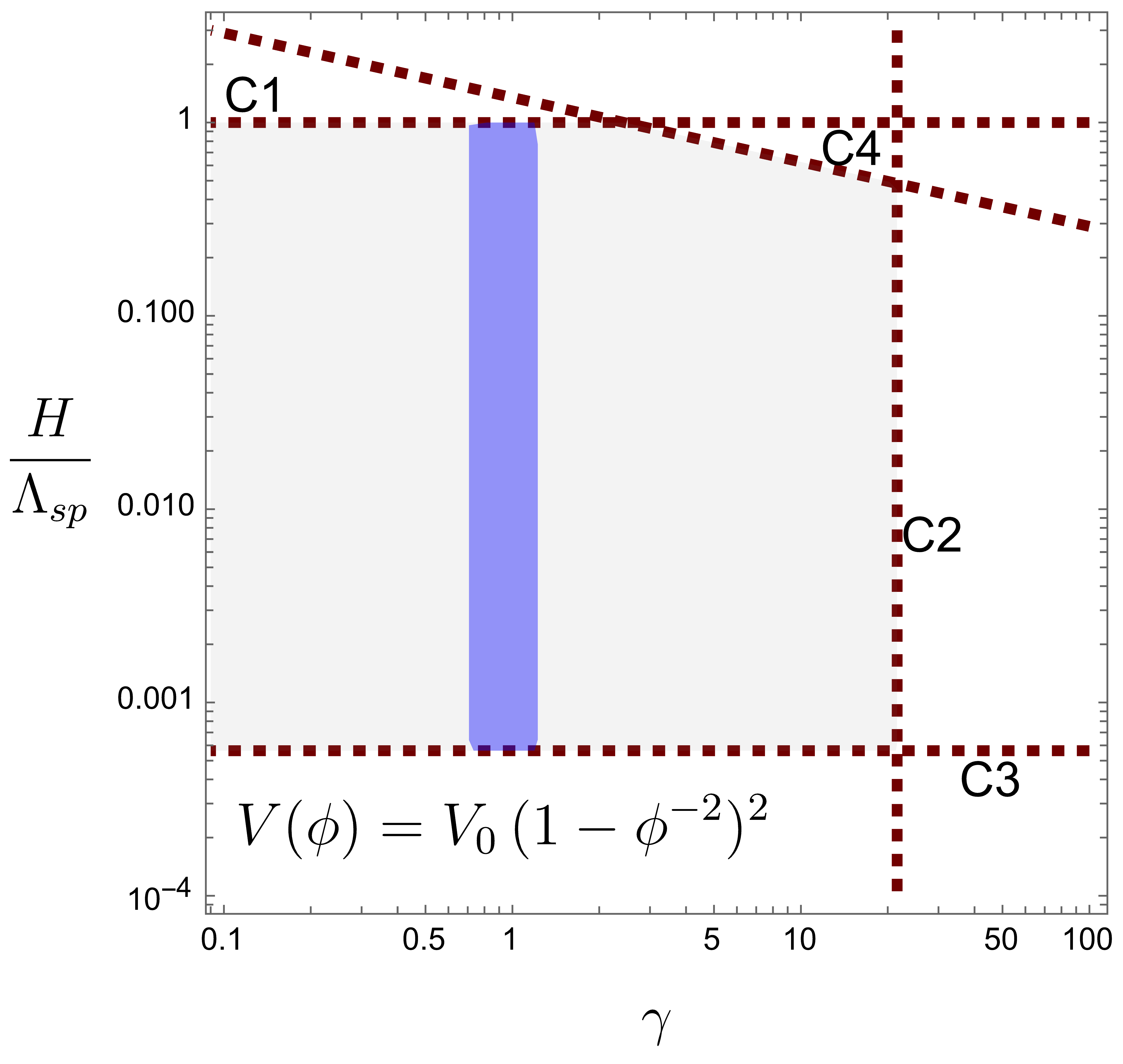}& \includegraphics[width=0.31\textwidth]{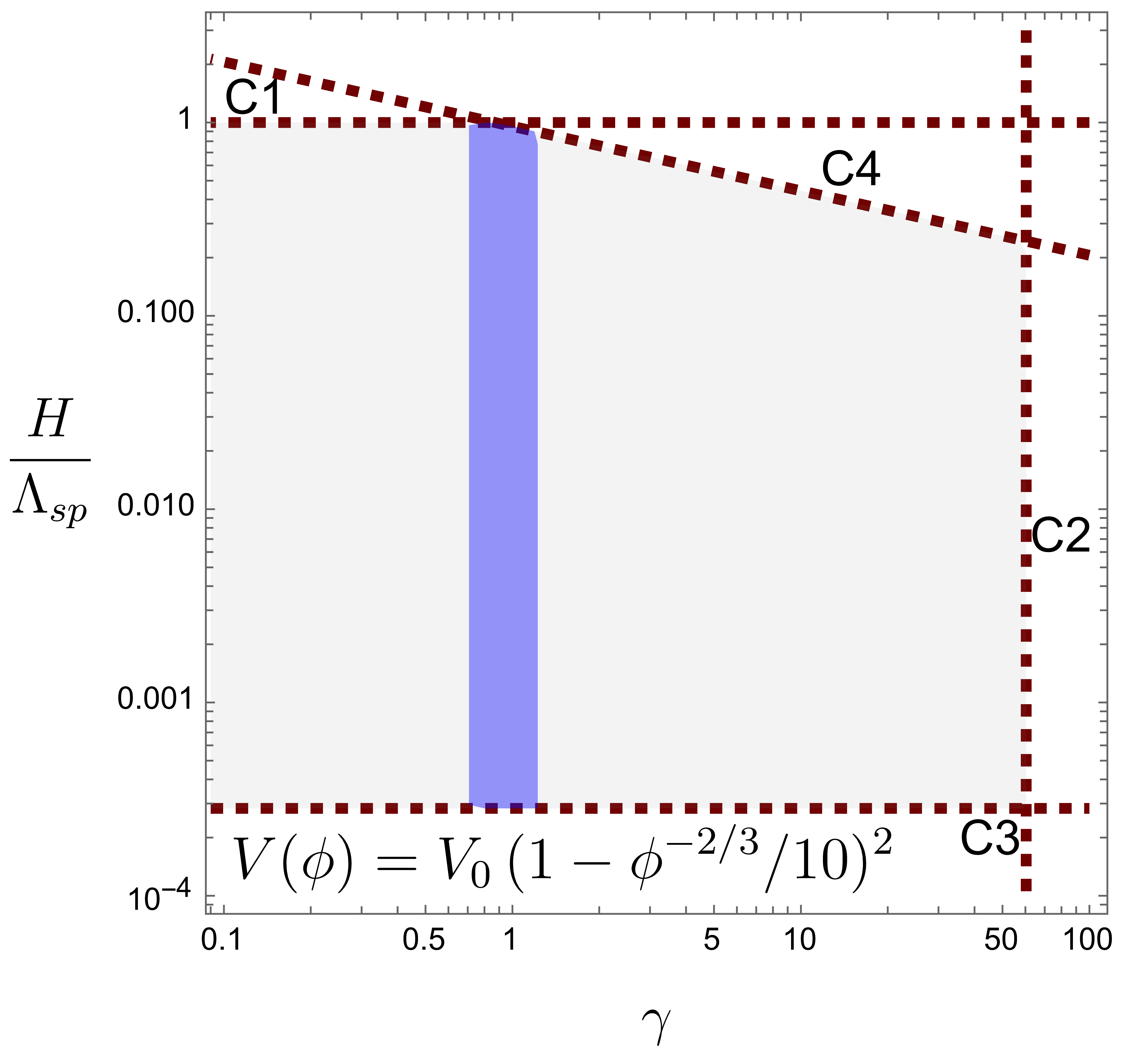}
        
    \end{tabular}
    \caption{Region allowed by the constraints imposed by the conditions reported in~\Cref{tab:constraints} (red dotted lines) for the parameter space of our model, in terms of the species coupling $\gamma$ vs the ratio between IR and UV scales $H/\Lambda_{\text{sp}}$. We show the regions for several example potentials considered in this work. The allowed regions are represented by shaded grey areas, with purple bands highlighting the values of $\gamma$ compatible with the SDC.}
    \label{fig:multiple_constraints}
\end{figure}

In~\Cref{fig:multiple_constraints}, we show all these constraints in the $\gamma-H/\Lambda_{\text{sp}}$ plane. Interestingly, sizable portions of the parameter space remain open after imposing $\gamma \sim \mathcal{O}(1)$ and $H/\Lambda_{\text{sp}}<1$. For reference, in each of these plots, we illustrate the typical values of the exponential coupling allowed by SDC (see~\cref{eq:sdconst}) with a purple band. As expected, these bands always lie inside the region compatible with our computational scheme and with the constraints we impose. Although in these plots we saturate the C1 condition by taking $H\simeq \Lambda_{\text{sp}}=1$, we reiterate that, at this scale, the effective description would completely break down due to QG effects. In fact, it has been argued that for $H / \Lambda_{\text{sp}} \simeq 1$, there should be a complete breakdown of the field theoretical formalism itself, as higher spin operators become relevant~\cite{Porrati:2008rm}. This is well known from a string theoretical point of view, but it has also been successfully argued from a bottom-up approach~\cite{Camanho:2014apa, Caron-Huot:2024lbf}. While condition C4 in~\cref{eq:field_velocity} was introduced to ensure that we can proceed perturbatively in $\gamma$, we can notice that for the potentials considered, it is trivially guaranteed by conditions C1 and C2, which are  physically more motivated. 

\begin{figure}[t]
		\centering
               \includegraphics[width=0.72\textwidth]{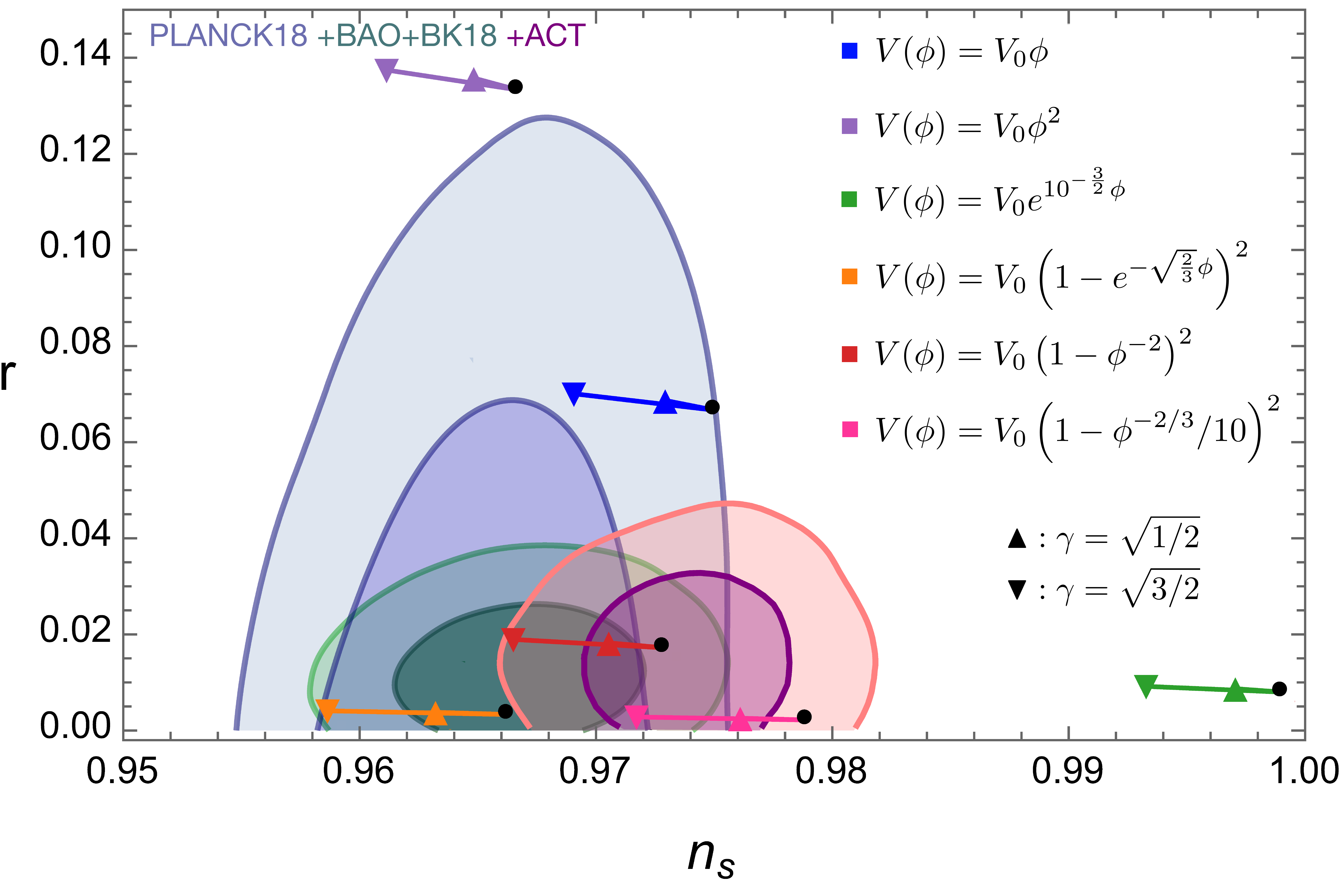}
		\caption{Theoretical predictions for $n_s$ vs $r$ all potentials considered in this work (see main legend) compared with observational constraints represented by colorful shaded areas (see top left legend and main text for the different data combinations).  We stress that any meaningful deviations from single-field inflation predictions (corresponding to $H/\Lambda_{\text{sp}}\simeq 0$ and represented with black dots) only occur near $H/\Lambda_\text{sp}\simeq$ 1 where the EFT description is expected to break down. To not saturate this bound, the maximal value we employ in this plot is $\Lambda_{\text{sp}}\leq 0.5$, where all colorful lines end in either upward ($\blacktriangle$) or downward ($\blacktriangledown$)  colorful triangles, representing $\gamma =\sqrt{1/2}$ and $\gamma =\sqrt{3/2}$, respectively. }
		\label{fig:testfig}
	\end{figure}

	In ~\Cref{fig:testfig}, we present the modified predictions for $n_s$ and $r$ for the same models shown in~\Cref{fig:multiple_constraints} and summarized in~\Cref{tab:modelstable} compared with the observational constraints imposed by Planck~\cite{Planck:2018vyg} (blue shaded area) + BAO + BICEP/Keck (BK18) data~\cite{BICEP:2021xfz} (green shaded area) + Atacama Cosmology Telescope (ACT)~\cite{ACT:2025tim} (shaded pink area). For each model, we show two reference points, going from $H/\Lambda_{\text{sp}} = 0$, where the corrections induced by the tower are negligible, to $ H/\Lambda_{\text{sp}} = 0.5$, where the corrections are sizable. For each model, we show the two reference values $\gamma = \sqrt{1/2}$ and $\gamma = \sqrt{3/2}$, which we highlight with different symbols (upwards/downwards triangles, respectively). Notice that for all models of our interest, these two curves are nearly overlapping, except for the fact that they terminate at different points of the parameter space. This signals that the corrections of the tower of light modes can, in fact, be neglected unless one approaches the energy scale at which the EFT breaks down. For all cases considered in this work, the tower-induced corrections tend to reduce the value of $n_s$ and to slightly increase the value of $r$ compared to the case where the tower is not present. We highlight that inverse Hilltop potentials, which can naturally arise in string constructions~\cite{Burgess:2001fx, Bansal:2024uzr}, provide the best agreement with the most recent observational data.

\section{Conclusions}
 In this work, we have analyzed the dynamics of inflation when coupled to an infinite tower of light scalar modes through their exponential mass term, as suggested by the Swampland Distance Conjecture (SDC) and as given by eq.~\eqref{eq:full_action_exp}. Specifically, we have studied whether the presence of the tower can have significant effects for the EFT at cosmological energy scales, such as developing nontrivial dynamical and observational implications due to their coupling to the inflaton $\varphi$, particularly through particle production. To analyze these effects in a controlled setup, we have considered a mass coupling of the form $m_{\text{t}} \sim e^{-\gamma \varphi}$, and studied its impact on inflationary observables. We computed the scalar and tensor power spectra, incorporating both single-field contributions and the additional sourced contributions arising from particle production of tower modes. We carried out the analysis in a fully time-dependent setting, which allowed us to derive analytic expressions for the corrections to the scalar spectral index $n_s$, the tensor-to-scalar ratio $r$, and the non-Gaussianity parameter $f_{\rm NL}$. These can be found in Sec.~\ref{sec:perturbations}. We found that all corrections scale universally as
\begin{equation}
\delta\{n_s, r, f_{\rm NL}\} \propto \left(\frac{H}{\Lambda_{\text{sp}}}\right)^{2+p} \; ,
\end{equation}
where $p \geq 1$ parametrizes the density of states in the tower, $H$ is the Hubble parameter during inflation and $\Lambda_{\rm sp}$ is the species scale, taken as gravitational cut-off of the inflationary EFT. Our results show that, provided that the EFT remains weakly coupled, the presence of towers of species during inflation does not directly alter observational predictions. That is, as long as $H \ll \Lambda_{\text{sp}}$, the corrections to observables remain parametrically small. The inflationary predictions are therefore robust under such couplings in the regime where the EFT remains valid. In fact, the species scale $\Lambda_{\text{sp}}$ sets a limit not just on the UV cutoff of the EFT but also on the degree to which towers of states can influence inflationary dynamics, and significant deviations from standard inflationary predictions can only arise when the inflationary dynamics approach the quantum gravity regime, \emph{i.e.}, when $H \sim \Lambda_{\text{sp}}$, where the EFT formalism breaks down. We illustrated these statements with explicit examples, considering models of large field inflation, including monomial, power-law, Starobinsky-like, and inverse hilltop potentials. In all cases, the corrections due to the towers were found to be negligible during inflation. This is also visually displayed in Fig.~\ref{fig:testfig}. 

In conclusion, we have shown that while towers of light states, as predicted by the SDC, introduce new degrees of freedom into the inflationary EFT, their impact on observational quantities remains marginal as long as the model stays within the EFT regime of validity.

A final remark concerns the KK interpretation of the light scalars. If these scalars are identified as spin-0 modes of a KK tower, the tower necessarily includes massive spin-2 modes (KK gravitons), which are constrained by the Higuchi bound. Since the mass scales of spin-0 and spin-2 modes are generically not parametrically decoupled, this bound would require the typical mass of the tower to lie above the Hubble scale. As discussed in Sec.~\ref{sec:heavymodes}, towers with masses above Hubble give exponentially suppressed contributions, rendering their effects negligible for inflationary dynamics.

To avoid this tension, one would need a mechanism generating different mass gaps for KK gravitons and scalars. Such hierarchies have been studied in linear dilaton backgrounds~\cite{Antoniadis:2021ilm}, where the graviton mass gap is smaller than the scalar one. Other studies have focused on warped compactifications, both with  scalars~\cite{Goldberger:1999uk,Goldberger:1999un,Grzadkowski:2003fx} and fermions~\cite{Megias:2019vdb}.

 In our setup, we do not specify a concrete mechanism, but rather assume that an effective scale separation arises because only the KK scalars couple directly to the inflaton. This coupling can suppress their masses relative to spin-2 modes, ensuring that the EFT remains consistent. A more detailed study of KK towers in positively curved backgrounds, and the implications for unitarity, remains an interesting direction for future research.

\label{sec:conclusions}
	\acknowledgments
	J.M. would like to thank the group at Insituto de Fisica Teórica (IFT) in Madrid for their helpful comments and discussion, some of which are reflected in the revised manuscript. We thank Angelo Ricciardone for his comments on the draft. We are especially grateful to Matteo Braglia for his insightful comments on a late version of this manuscript. We would also like to thank Cumrun Vafa for his comments on an earlier version of this draft. The work of M.P. is supported by the Comunidad de Madrid under the Programa de Atracción de Talento Investigador with number 2024-T1TEC-3134. M.P. acknowledges the hospitality of Imperial College London, which provided office space during some parts of this project. 

	\appendix
    \crefalias{section}{Appendix}
 
	\section{Towers with non-linear scaling}
   \label{sec:appendix_powers}
     We have only discussed towers with a linear mass spectrum; here, we generalize our results to towers with general monomial mass scaling as
	\begin{equation}
		m_n=m_{\text{t}} n^{1/p} \;  ,
	\end{equation}
	with $p>1$. The number of species $N$ is defined in terms of the QG cutoff as
	\begin{equation}
		\Lambda_{\text{sp}}=\dfrac{M_{\rm P}}{\sqrt{N}} \; , 
	\end{equation}
	\begin{equation}
		m_{N}=N^{1/p}m_{\text{t}}\simeq \Lambda_{\text{sp}} \; .
	\end{equation}
	This scale is related to the mass of the tower as
	\begin{equation}
		\Lambda_{\text{sp}}=M_{\rm P}^{\frac{2}{2+p}}m_{\text{t}}^{\frac{p}{2+p}} \;  .
	\end{equation}
	The number of states entering the inflationary dynamics can be similarly computed.
	\begin{equation}
		N_H\simeq\left(\dfrac{H}{m_{\text{t}}}\right)^{p} \; .
	\end{equation}
	Generalizing the results for $p>1$ is straightforward; we compute the field derivative as
	\begin{equation}
		\dot\varphi=-\dfrac{V'(\varphi_0)}{3H}\left(1-\dfrac{\gamma}{M_{\rm P}}\dfrac{3}{16 \pi ^2}\dfrac{V}{V'}\left(\dfrac{3}{2}\right)^{p-1}\left(\dfrac{H}{\Lambda_{\text{sp}}}\right)^{2+p} \right)\; .
		\label{eq:phidotp1}
	\end{equation}
	This means that the single-field theory receives corrections of the order
	\begin{equation}
		\left(\dfrac{H}{\Lambda_{\text{sp}}}\right)^{2+p}
	\end{equation}
	We can also compute the corresponding expressions for the relevant inflationary quantities.
	\begin{equation}
		P_\zeta(k)=\dfrac{H^4}{(2\pi)^2\dot{\varphi}^2} \left[1+ \frac{1}{8 \pi^2}  \dfrac{ \gamma^2   }{ 27   }\left(\dfrac{3}{2}\right)^{p-1}\dfrac{3p}{2+p}\left(\dfrac{H}{\Lambda_{\text{sp}}}\right)^{2 + p}\right]\; .
		\label{eq:pzetafinalp}
	\end{equation}
	\begin{equation}
		n_s \simeq 1 +2\eta  - 4 \ve  +   \frac{\gamma^2}{72 \pi^2}  \left(\dfrac{3}{2}\right)^{p-1}\dfrac{3p}{2+p} \left(\dfrac{H}{\Lambda_{\text{sp}}}\right)^{2 + p}  \ve \;  ,
	\end{equation}
	\begin{equation}
		r\simeq 16\epsilon\,\left[1-\dfrac{\gamma}{M_{\rm P}}\dfrac{3}{16 \pi ^2}\dfrac{1}{\sqrt{2\epsilon}}\left(\dfrac{3}{2}\right)^{p-1}\left(\dfrac{H}{\Lambda_{\text{sp}}}\right)^{2 + p} \right]
		\; ,
	\end{equation}
	\begin{equation}
		f^{\rm equil}_{\rm NL}=  \dfrac{\gamma^3 }{2916 \pi^2} \,   \sqrt{ 2 \ve } \,  \dfrac{5p}{4+p}\left(\dfrac{H}{\Lambda_{\text{sp}}}\right)^{2 + p} \; .
	\end{equation}
	
	\section{Correlation functions between heavy tower modes}
	\label{sec:appendix_modes}
    In the main text, we have mainly focused on light modes with masses below the Hubble scale. Here, we justify this by showing that the heavier modes produce a contribution which is parametrically the same, but highly suppressed by numerical factors.
    In the small $k \tau$ limit, the solution for the heavy takes the form
	\begin{equation}
		\begin{aligned}
			\xi_n(\tau, \vec{k}) = \sqrt{-\tau}e^{-\frac{1}{2} (\pi  \sqrt{\delta_n} )} & \left\{  \frac{\sqrt{\pi } \, e^{\frac{i \pi }{4}} 2^{-1-i \sqrt{\delta_n} }   \left[ \coth (\pi  \sqrt{\delta_n} )+1\right]}{ \Gamma (i \sqrt{\delta_n} +1)}(-k\tau)^{i\sqrt{\delta_n}} + \right. \\
			& \hspace{.4cm} \left.  + \frac{e^{-i\frac{\pi}{4}} 2^{-1+i \sqrt{\delta_n} } \Gamma (i \sqrt{\delta_n} )}{\sqrt{\pi }}(-k\tau)^{-i\sqrt{\delta_n}} \right\}.
		\end{aligned}
	\end{equation}
The modes can be expressed as
\begin{equation}
 \xi_{n} ( \tau,\vec{k}) = \frac{1}{\sqrt{2 \, \omega_n \left( \tau \right)}} \left[ 
 \alpha_n \left( \tau, k \right) \, {\rm e}^{-i \int^\tau d \tau' \omega_n \left( \tau' \right) } + 
  \beta_n \left( \tau, k \right) \, {\rm e}^{i \int^\tau d \tau' \omega_n \left( \tau' \right) } \right] \;, 
\label{boloyubov}
\end{equation} 
with 
\begin{equation}
    \omega_n=\dfrac{\sqrt{\delta_n}}{-\tau}.
\end{equation}
This allows us to identify the Bogoliubov coefficients~\cite{Kofman:1997yn} as
	\begin{eqnarray}
	\alpha_n & = & \delta_n^{1/4} e^{-\frac{1}{2} (\pi  \sqrt{\delta_n} )} \frac{\sqrt{\pi} e^{\frac{i \pi }{4}} 2^{-1/2-i \sqrt{\delta_n} }  \left[\coth (\pi  \sqrt{\delta_n} )+1\right]}{  \Gamma (i \sqrt{\delta_n} +1)} \; , \\
		\beta_n & = & \delta_n^{1/4} e^{-\frac{1}{2} (\pi  \sqrt{\delta_n} )}\frac{e^{-i\frac{\pi}{4}} 2^{-1/2+i \sqrt{\delta_n} } \Gamma (i \sqrt{\delta_n} )}{\sqrt{\pi }} \; ,
	\end{eqnarray}
	such that $|\alpha|^2-|\beta|^2=1$. Then, we can express the two-point correlation function in momentum space as
	\begin{equation}
		\langle\chi_n(\tau, \vec{k})\text{ } \chi_n(\tau, \vec{k'})\rangle=\dfrac{1}{a^2}\delta(\vec{k}+\vec{k'})\dfrac{\tau}{\sqrt{\delta_n}}|\beta_n|^2 \; .
	\end{equation}
	In terms of $\vec{x}$ this reads 
	\begin{equation}
		\langle \chi_n(\tau, \vec{x})\text{ } \chi_n(\tau, \vec{x})\rangle=\dfrac{1}{a^2}\int_{0}^{\frac{\sqrt{\delta_n}}{\tau}} \dfrac{4\pi k^2 \textrm{d} k}{(2\pi)^3}\dfrac{\tau}{\sqrt{\delta_n}}|\beta_n|^2 =\dfrac{1}{a^2}\frac{e^{-2 \pi  \sqrt{\delta_n} }\delta_n}{6 \pi ^2 \tau ^2}=\frac{e^{-2 \pi  n/N_H }m_1^2e^{-2\gamma \vp} n^2}{6 \pi ^2} \; ,
	\end{equation}
	where he have used~\cref{eq:NH} and~\cref{eq:scalingofmasses}. As we can see, the heavy mode contribution is exponentially suppressed as the mass increases.
	
	To compute the total contribution from heavy modes, we should sum over all the heavy modes up to the QG cutoff. For this purpose, it is useful to note that, using
    \begin{equation}
		\sum_{k=N_1+1}^\infty k^n z^k=\left(z\dfrac{\partial}{\partial z}\right)^n \dfrac{z^{N_1}}{1-z} \; ,
	\end{equation}
    we can approximate the sum 
    \begin{equation}
		\sum_{n=3 N_H/2+1}^N n^4e^{-2\pi n/N_H}\simeq 10^{-4}N_H^5 \; ,
	\end{equation}
    where we have assumed $1\ll N_H \ll N$. Using these, the sum over all heavy modes can be expressed as
	\begin{equation}
		\sum_{n=3N_H/2+1}^N\dfrac{m_n^2}{2} e^{-2\gamma \varphi}
		\langle\chi_n(\tau, \vec{x})\ \chi_n(\tau, \vec{x})\rangle\simeq\dfrac{4\times10^{-5}}{\pi^2}H^4 N_H
		\label{eq:twopcorrexph} \; .
	\end{equation}
	Comparing~\cref{eq:NH_Lambda_eq} with~\cref{eq:twopcorrexph}, we can see that the heavy modes contribute negligibly compared to the light modes.

    In the case in which there are no light modes below the Hubble scale we have
        \begin{equation}
        m_t\gg H,\quad N_H\ll1,
    \end{equation}
    and the sum is dominated by the first term
    \begin{equation}
		\sum_{n=1}^N\dfrac{m_n^2}{2} e^{-2\gamma \varphi}
		\langle\chi_n(\tau, \vec{x})\ \chi_n(\tau, \vec{x})\rangle\simeq\dfrac{e^{-2\pi m_t/H}}{12\pi^2} m_t^4
		\label{eq:twopcorrexph2b} \; .
	\end{equation}
    Expressed in terms of the species scale $\Lambda_{\mathrm{sp}}= m_t^{1/3}$ we have
        \begin{equation}
		\sum_{n=1}^N\dfrac{m_n^2}{2} e^{-2\gamma \varphi}
		\langle\chi_n(\tau, \vec{x})\text{ } \chi_n(\tau, \vec{x})\rangle\simeq\dfrac{e^{-2\pi \left(\frac{\Lambda_{\text{sp}}}{H}\right)^3 H^2}}{12\pi^2} \Lambda_{\text{sp}}^{12}
		\label{eq:twopcorrexph2c} \; .
	\end{equation}
    Then, we see that in the large $m_t$ limit the contribution of the modes goes decays exponentially fast towards 0, as we would expect from super Hubble modes. As the mass of the modes approaches the Hubble scale we instead find that the contribution is upper bounded as 
        \begin{equation}
		\sum_{n=1}^N\dfrac{m_n^2}{2} e^{-2\gamma \varphi}
		\langle\chi_n(\tau, \vec{x})\text{ } \chi_n(\tau, \vec{x})\rangle\lesssim H^4 \; .
	\end{equation}
If we compare this to \cref{eq:twopcorrexp}, we can conclude that the contribution of a tower coupled exponentially to the inflaton,  with mass gap starting at or above the Hubble scale, will have a negligible contribution during inflation.
	\section{Two- and three-point correlation functions}
	\label{sec:corrfuncs}
 In this appendix, we report the main ingredients for the derivation of the two and three-point correlation functions for scalar and tensor perturbations. Let us start by focusing on the scalar perturbations. The Green function $G(\tau,\tau')$ for the e.o.m. for scalar perturbations in~\cref{eq:sourcedeq} is given by
	\begin{equation}
		G(\tau,\tau')=
		\frac{\pi \sqrt{\tau  \tau '} }{4}   \theta \left(\tau -\tau '\right)\left[ Y_{\Delta}(k \tau ) J_{\Delta}\left(k \tau '\right)- J_{\Delta}(k \tau ) Y_{\Delta}\left(k \tau '\right)\right]\; ,
	\end{equation}
	where $\Delta \equiv \frac{1}{2} \sqrt{9-4\beta^2}$  and $J_{\Delta}$, $Y_{\Delta}$ are the Bessel functions of the first and second kinds. The sourced solution can be expressed as 
	\begin{equation}
		\psi_s (\tau, \vec{k})=\int^\tau_{
			-\infty} \textrm{d} \tau' G(\tau,\tau')S(\tau',\vec{k})\; ,
	\end{equation}
	where $S(\tau,\vec{k})$ is the source term in~\cref{eq:sourcedeq} given by
	\begin{equation}
		\begin{aligned} 
			S(\tau,\vec{k}) & = \sum_n \dfrac{\gamma \, \delta_n (\tau) }{ H (-\tau)^3}  (\chi_n\chi_n-\langle\chi_n\chi_n\rangle) \\ 
			& =  \sum_n \dfrac{\gamma \, \delta_n (\tau)}{  H (-\tau)^3} \left[  \int\dfrac{\textrm{d} \vec{p} }{(2\pi)^{3/2}}\chi_{n,\vec{p}}\hspace{1mm} \chi_{n,\vec{k}-\vec{p}} - \langle \chi_n\chi_n \rangle \right] \; .
		\end{aligned}
	\end{equation}
	The two-point correlation function is given by
	\begin{equation}
		\label{eq:two_point_intermediate}
		\langle \psi_s(\tau, \vec{k}_1) \psi_s(\tau, \vec{k}_2)\rangle=\int^{\tau}_{
			-\infty} \textrm{d} \tau_1 G(\tau,\tau_1)\int^{\tau}_{
			-\infty} \textrm{d} \tau_2 G(\tau,\tau_2)\langle S(\tau_1,\vec{k}_1)S(\tau_2,\vec{k}_2)\rangle  \; .
	\end{equation}
	The expectation value of the source term can be expressed as
	\begin{equation}
		\langle S(\tau_1,\vec{k}_1)S(\tau_2,\vec{k}_2)\rangle=\sum_{n, m} \dfrac{\gamma^2 \, \delta_n (\tau_1) \delta_m (\tau_2) }{  H^2 (-\tau_1)^3 (-\tau_2)^3 } \left[ \int\dfrac{\textrm{d} \vec{p} \, \textrm{d} \vec{q}}{(2\pi)^{3}}\langle\chi_{n,\vec{p}}\hspace{0.5mm} \chi_{n,\vec{k}_1-\vec{p}}\chi_{m,\vec{q}}\hspace{0.5mm} \chi_{m,\vec{k}_2-\vec{q}}\rangle - \delta_{nm} (\langle \chi_n \chi_n \rangle )^2 \right] \; ,
	\end{equation}
	which, assuming the fields to be Gaussian, reduces to 
	\begin{equation}
		\langle S(\tau_1,\vec{k}_1)S(\tau_2,\vec{k}_2)\rangle=
		\sum_{n, m} \dfrac{2 \gamma^2 \, \delta_n (\tau_1) \delta_m (\tau_2)  }{ H^2 (-\tau_1)^3 (-\tau_2)^3 } \int \dfrac{\textrm{d} \vec{p} \, \textrm{d} \vec{q}}{(2\pi)^{3}} \langle\chi_{n,\vec{p}}\hspace{0.5mm} \chi_{m,\vec{k}_2-\vec{q}}\rangle \langle\chi_{m,\vec{q}}\hspace{0.5mm}\chi_{n,\vec{k}_1-\vec{p}}\rangle \; . 
	\end{equation}
	Notice the factor 2 that comes from the sum over all possible contractions. We proceed by substituting~\cref{eq:chi_correlation_full} and integrating over $\vec{q}$ to get
\begin{equation}
	\label{eq:source_intermediate}
	\langle S(\tau_1,\vec{k}_1)S(\tau_2,\vec{k}_2)\rangle =   \sum_{n} \dfrac{2 \gamma^2 \, \delta_n (\tau_1) \delta_n (\tau_2) H^2  }{ (-\tau_1) (-\tau_2) } \delta(\vec{k}_1+\vec{k}_2)\int\dfrac{\textrm{d} \vec{p}}{(2\pi)^{3}}|\xi_n(\tau_1,\vec{p})|^2   |\xi_n(\tau_2,\vec{p}+\vec{k}_2)|^2 \; .
\end{equation}
The $\vec{p}$ integral, say $\mathcal{I}$ can be further expanded as
\begin{equation}
	\begin{aligned}
		\mathcal{I} & \simeq \int_{0}^{\frac{\sqrt{2-\delta_n}}{ \tau_{\rm M} }}\dfrac{2\pi p^2 \,\sin{\theta} \, \textrm{d}p \, \textrm{d}\theta }{(2\pi)^{3}}\frac{\left(p^2 \tau_{1}^{2}\right)^{\delta_n /3} \left(\tau_{2}^{2} \left(k^2+2 k p \cos\theta +p^2\right)\right)^{\delta_n /3}}{ 4 p^3 \tau_{1}^{2} \tau_{2}^{2} \left(k^2+2 k p \cos\theta +p^2\right)^{3/2}} \\
		& \simeq\dfrac{3 \left(k^2 \tau_1 \tau_2\right)^{\delta_n  /3}}{16 \pi ^2 \delta_n (\tau_{\rm M}) k^3 \tau_1^2\tau_2^2} \left(\dfrac{(2-\delta_n) \tau_1\tau_2}{ \tau_{\rm M}^2 }\right)^{\delta_n  /3}\simeq\dfrac{3}{16 \pi ^2 \delta_n  k^3 \tau_1^2\tau_2^2} \; , 
	\end{aligned}
\end{equation}
where in the first line we have used the small $\tau_1$, $\tau_2$ solutions for $\xi_n$ and in the second line we have defined $-\tau_{\rm M} \equiv \text{Max}(-\tau_1, -\tau_2) $, expanded at small $p$ and used $\delta_n \ll 1$. Substituting into~\cref{eq:source_intermediate} we get
\begin{equation}
	\label{eq:source_2p_final}
	\langle S(\tau_1,\vec{k}_1)S(\tau_2,\vec{k}_2)\rangle =   \sum_{n} \dfrac{ 3 \gamma^2  H^2  }{8 \pi ^2   k^3 (-\tau_1)^3 (-\tau_2)^3 } \frac{\delta_n (\tau_1) \delta_n (\tau_2) }{\delta_n(\tau_{\rm M})}  \delta(\vec{k}_1+\vec{k}_2)\; .
\end{equation}
We notice that the two integrals in $\tau_1$, $\tau_2$ in~\cref{eq:two_point_intermediate} factorize. We can thus compute
\begin{equation}
	\int^\tau_{
		-\infty} \textrm{d} \tau' \dfrac{1}{\tau'^3}G(\tau,\tau') \simeq - \frac{1 }{9 \tau} \; , 
\end{equation}
where we have expanded the Green function using $\Delta \simeq \frac{3}{2} - \frac{\beta^2}{3}$. Substituting~\cref{eq:source_2p_final} and the two integrals into~\cref{eq:two_point_intermediate} we get
\begin{equation}
	\langle \psi_s(\tau, \vec{k}_1) \psi_s(\tau, \vec{k}_2)\rangle=  \sum_{n} \dfrac{ \gamma^2  H^2  }{8 \pi ^2   k^3   } \frac{\delta_n (\tau)  }{27 \tau^2} \; \delta(\vec{k}_1+\vec{k}_2) \; . 
\end{equation}
The sourced power spectrum thus reads
\begin{equation}
	P_\zeta^{\chi} = \frac{k^3}{2 \pi^2} \frac{H^2}{\dot \vp_0^2  }  \frac{1}{a^2 } \langle \psi_s(\tau, \vec{k}_1) \psi_s(\tau, \vec{k}_2)\rangle = \frac{1}{16 \pi^4} \frac{H^4}{\dot \vp_0^2  }  \dfrac{ \gamma^2  H^2  }{ 27    } \sum_{n}  \delta_n (\tau)  \; .
\end{equation}
Finally, we get
\begin{equation}
	\label{eq:sourced_spectrum_final}
	P_\zeta^{\chi} = \frac{1}{16 \pi^4} \frac{H^4}{\dot \vp_0^2  }  \dfrac{ \gamma^2   }{ 81 }  N_H H  \; ,
\end{equation}
where we have used~\cref{eq:scalingofmasses} and~\cref{eq:NH}, leading to $\sum_n \delta_n(\tau) = N_H / 3$, to compute the sum over all the modes in the tower.

Following a similar approach, we estimate the three-point correlation function for scalar perturbations
\begin{equation}
	\begin{aligned}
		\langle \psi (\tau, \vec{k}_1) \psi (\tau, \vec{k}_2) \psi (\tau, \vec{k}_3)\rangle & = \int^{\tau}_{
			-\infty} \textrm{d}\tau_1 G(\tau,\tau_1)\int^{\tau}_{
			-\infty} \textrm{d}\tau_2 G(\tau,\tau_2) \; \times \\
		& \qquad \times \int^{\tau}_{
			-\infty} \textrm{d} \tau_3 G(\tau,\tau_3)\langle S(\tau_1,\vec{k}_1)S(\tau_2,\vec{k}_2)S(\tau_3,\vec{k}_3)\rangle \; .
	\end{aligned}
\end{equation}
For this purpose, we start by computing
\begin{equation}
	\begin{aligned}
		\langle S(\tau_1,\vec{k}_1)S(\tau_2,\vec{k}_2)S(\tau_3,\vec{k}_3)\rangle  = & \, 8  \gamma^3 H^6 \sum_{m, n, l}  a(\tau_1)^3 a(\tau_2)^3a(\tau_3)^3 \delta_{m} \delta_{n} \delta_{l} \; \times \\
		\times \; \int\dfrac{\textrm{d}\vec{p} \,  \textrm{d} \vec{q} \, \textrm{d} \vec{s} }{(2\pi)^{9/2}} \, &  \langle\chi_{m,\vec{p}}\, \chi_{n,\vec{q}} \rangle \, \langle \chi_{n,\vec{k}_2-\vec{q}} \chi_{l,\vec{s}} \rangle \, \langle \chi_{m,\vec{k}_1-\vec{p}}  \chi_{l,\vec{k}_3-\vec{s}}\rangle \\
		= & \, 8 \gamma^3 H^6 \sum_{ n }  a(\tau_1) a(\tau_2)a(\tau_3) \delta_{n}(\tau_1)  \delta_{n} (\tau_2) \delta_{n} (\tau_3)  \; \times \\
		\delta(\vec{k}_1+\vec{k}_2+\vec{k}_2) \,  \int\dfrac{\textrm{d}\vec{p} }{(2\pi)^{9/2}} & \, |\xi_n(\tau'_1,\vec{p})|^2   |\xi_n(\tau'_2,\vec{p}+\vec{k}_2)|^2|\xi_n(\tau'_3,\vec{p}-\vec{k}_1)|^2 \; ,
	\end{aligned}
\end{equation}
where the 8 is the combinatorial factor keeping track of all possible Wick contractions.  We expect the bispectrum to be maximal for equilateral configurations, as is common for particle production, so we restrict ourselves to the case $k_1=k_2=k_3$, and we substitute~\cref{eq:solxi} to get
\begin{equation}
	\langle S(\tau_1,\vec{k})S(\tau_2,\vec{k})S(\tau_3,\vec{k})\rangle\simeq  8 \gamma^3 H^3 \sum_{ n }   \frac{ \delta_{n}(\tau_1)  \delta_{n} (\tau_2) \delta_{n} (\tau_3) }{\delta_n (\tau_M) } \dfrac{1}{\tau_1^3\tau_2^3\tau_3^3}\dfrac{3}{64\sqrt{2}\pi^{7/2}k^6}
\end{equation}
The three-point correlation function is then:
\begin{equation}
	\langle \psi (\tau, \vec{k}) \psi (\tau, \vec{k})\psi (\tau, \vec{k})\rangle=  \left( \frac{2 \gamma H }{9 \tau } \right)^3 \dfrac{3}{64\sqrt{2}\pi^{7/2}k^6} \frac{N_H}{5} \; ,
	\label{eq:three_point_final}
\end{equation}
where we have used $\sum_n \delta_n^2 = N_H / 5$.

Finally, we focus on tensor perturbations. We start by considering the Green function $G(\tau,\tau')$ for the tensor e.o.m. in~\cref{eq:eom_tensors}, given by
\begin{align}
	G_k(\tau,\tau')&=\frac{\Theta\left(\tau-\tau'\right)}{k^3\,\tau \tau'}\Big[\left(1+k^2\,\tau\,\tau'\right)\sin k\left(\tau-\tau'\right) \,+\, k\left(\tau'-\tau\right) \,\cos k\left(\tau-\tau'\right)\Big]\,.
\end{align}
The sourced solutions can then be computed as
\begin{equation}\label{eq:tenssol}
	h_{ij}(\vec{k},\,\tau)=2 \int \textrm{d} \tau' G_k(\tau,\,\tau')\,\Pi_{ij}{}^{ab}( \vec{k} )\,T_{ab}(\vec{k},\,\tau')\; .
\end{equation}
First of all, we notice that
\begin{equation}
    \Pi_{ij}{}^{ab} \Pi_{ij}{}^{cd} \, p_a \,  p_b  \,p_c  \,p_d = \dfrac{1}{2} \left[ \frac{(\vec{p} \cdot \vec{k} )^2}{k^2} - p^2 \right]^2    \; ,
\end{equation}
and, that the only part of $T_{ab}$ contributing to $ \Pi_{ij}{}^{ab} T_{ab}$ is the one proportional to $ \partial_a \chi_i \partial_b \chi_i $. Moreover, assuming the $\chi_i$ fields to be Gaussian, the two-point correlation function of the tensor perturbations reduces to
\begin{equation}
	\begin{split}
		\langle h_{ij}(\vec{k}_1,\,\tau)  h_{ij}(\vec{k}_2,\,\tau)\rangle= 2 \; 
		\delta(\vec{k}_1+\vec{k}_2) \int_{-\infty}^{\tau} \textrm{d}\tau_1 \; \frac{G_{k_1}(\tau,\tau_1)}{a^2(\tau_1)}  \;   \int_{-\infty}^{\tau} \textrm{d}\tau_2  \;  \frac{G_{k_1}(\tau,\tau_2) }{a^2(\tau_2)} \times \\
  \times \sum_n \int \dfrac{ \textrm{d} \vec{p}  }{(2\pi)^3 } \; \left[ \frac{(\vec{p} \cdot \vec{k}_1 )^2}{{k_1}^2} - p^2 \right]^2 \;  
		  | \xi_n(\tau_1,\vec{p})|^2   |\xi_n(\tau_2,\vec{p}+\vec{k}_1)|^2 \; .
	\end{split}
 \label{eq:hh_expression}
\end{equation}
We proceed by using~\cref{eq:solxi} and noticing that the dominant contribution to the integral over internal momentum (corresponding to the integral in the second line of~\cref{eq:hh_expression}), say $\mathcal{J}$, comes from the $p>>k_1$ limit, to get
\begin{equation}
\label{eq:J_def}
\begin{aligned}
\mathcal{J}&=  \tau_1^{- 2 + 2 \delta_n/3}  \tau_2^{ - 2 +  2 \delta_n/3} \,
 \int_0^{\frac{\sqrt{2-\delta_n}}{\tau_M }} 
 \dfrac{2\pi p^2 \,\sin{\theta}  \, \textrm{d}\theta \, \textrm{d}p }{4 (2\pi)^{3}} \; p^4 \sin^4 \theta  \; 
  \frac{ p^{2 \delta_n /3} \left(  k_1^2+2 k_1 p \cos\theta +p^2\right)^{\delta_n /3}}{ p^3  \left(k_1^2+2 k_1 p \cos\theta +p^2\right)^{3/2}} \; ,\\
  & \simeq \tau_1^{- 2 + 2 \delta_n/3}  \tau_2^{ - 2 +  2 \delta_n/3} \frac{\sqrt{2-\delta_n } }{5 \pi ^2 (3 + 4 \delta_n  ) \tau_M} \;  \left(\frac{2-\delta_n }{\tau_M ^2}\right)^{\frac{2 \delta_n }{3}} \; .
  \end{aligned}
\end{equation}
Without loss of generality, let us proceed by assuming $\tau_M=\tau_1$ so that the double integral over $\tau_1$, $\tau_2$ reduces to
\begin{equation}
    \label{eq:tau_integral}
	\int \textrm{d} \tau_1 \, \textrm{d} \tau_2 \, G_{k_1}(\tau,\tau_1) \, \tau_1^{-1-2\delta_n/3} \,  G_{k_1}(\tau,\tau_2) \, \tau_2^{2\delta_n/3} \, \simeq \frac{9 \pi }{ 2 (6-4 \delta_n ) (3 + \delta_n)  k_1^3} \; .
\end{equation}
Substituting~\cref{eq:J_def} and~\cref{eq:tau_integral} into~\cref{eq:hh_expression}, expanding at the lowest order in $\delta_n$, and using the definition of the spectrum we get
\begin{equation}
	P_T^{\chi} (k) \simeq \dfrac{2\,H^2}{\pi^2 }\, \dfrac{H^2}{15 \sqrt{2} \pi} \sum_n  (1-0.79\delta_n ) \; .
\end{equation}
Finally, we can resolve the sum of the number of modes to get
\begin{equation}
	P_T^{\chi} (k)  \simeq \dfrac{.74}{15 \sqrt{2} \pi} \, \times  \dfrac{2\,H^2}{\pi^2 }\, N_H H^2 \simeq 1
 .1 \times 10^{-2} \, \times \dfrac{2\,H^2}{\pi^2 }\,  \left(N_H H^2  \right) \; 
 ,
\end{equation}
Finally, the tensor power spectrum, including the single-field contribution, reads
\begin{equation}
	\label{eq:tensor_spectrum_app}
	P_T^{\chi} \simeq \dfrac{2\,H^2}{\pi^2\,M_{\rm P}^2}\,\left[1+1.1 \, \times\,10^{-2} \, \times \left( \frac{ N_H H^2 }{\, M_{\rm P}^2} \right) \right] \; , 
\end{equation}
and, at the lowest order, the tensor-to-scalar ratio is
\begin{equation}
	r\simeq16\ve\left(1-\dfrac{1}{\sqrt{2\ve}}\dfrac{\gamma}{4\pi^2} N_H H^2 \right)
	\label{eq:reqexpf}\; .
\end{equation}
The sourced contribution to the tensor-to-scalar ratio will remain small in the parameter space analyzed, this is in agreement with the results of~\cite{Cook:2011hg, Dufaux:2007pt}, where it was shown that the presence of an adiabatically evolving gas of scalar particles does not contribute significantly to the tensor power spectrum and correspondingly to the tensor-to-scalar ratio.

\section{Large field inflation potentials}
\label{sec:appendix_models}
 For each model, we report the form of the potentials and the corresponding values of $\alpha$ and $\beta_\alpha$ for the parameterization in~\cref{eq:ve_universal}. Moreover, provide the modified expressions for the observable quantities $n_s$ and $r$ that are considered in this work.

\subsubsection*{Monomial inflation}

In general, for a monomial potential, we have
\begin{equation*}
V(\vp)=V_0\, \vp^{p},\quad\alpha=1,\quad\beta_\alpha=\dfrac{p}{4} \; ,
\end{equation*}
and we can express $n_s$ and $r$ as
\begin{equation}
    n_s-1=-\dfrac{p+2}{2 N_e}-\dfrac{3 \gamma}{16\pi^2}\sqrt{\dfrac{p}{2 N_e}}\left(\dfrac{H}{\Lambda_{\text{sp}}}\right)^3 \;,
\end{equation}
\begin{equation}
    r=\dfrac{4 p}{N_e}\left[1-\sqrt{\dfrac{2 N_e}{p}}\dfrac{3 \gamma}{16\pi^2}\left(\dfrac{H}{\Lambda_{\text{sp}}}\right)^3\right] \;.
\end{equation}

\subsubsection*{Power law inflation}

Exponential potentials of the form 
\begin{equation*}
V(\vp)=V_0\, e^{\lambda\vp},\quad\alpha=0\,\quad\beta_\alpha=\dfrac{\lambda^2}{2} \;,
\end{equation*}
lead to fixed $\ve>0$, and to a scale factor $a(t)$ that behaves not exponentially but as a power law over time. Since they have a vanishing second slow-roll parameter $\eta=0$, they lead to a relation $n_s-1=-r/8$, incompatible with observation. Coupled to a tower of states, they take the form
\begin{equation}
    n_s-1=-\lambda^2-\dfrac{3\gamma\lambda}{16\pi^2}\left(\dfrac{H}{\Lambda_{\text{sp}}}\right)^3 \;,
\end{equation}
\begin{equation}
    r=8\lambda^2\left[1-\dfrac{6}{\lambda}\dfrac{\gamma}{16\pi^2}\left(\dfrac{H}{\Lambda_{\text{sp}}}\right)^3\right] \;.
\end{equation}
For the analysis considered in this work, we fix $\lambda=10^{-3/2}$ as a benchmark.
    
\subsubsection*{Starobinsky inflation}

	The Starobinsky model arises from $R^2$ corrections to the Einstein-Hilbert action. The additional degrees of freedom can be expressed in terms of a canonically normalized scalar field with a potential of the form
	\begin{equation}
V(\vp)=V_0\left(1-e^{-\lambda\phi}\right)^2,\quad\alpha=2,\quad\beta_\alpha=\dfrac{1}{2 \lambda^2} \;.
	\end{equation}
 where canonically $\lambda=\sqrt{2/3}$. We have
\begin{equation}
    n_s-1\simeq-\dfrac{2}{N_e}-\dfrac{3\gamma/\lambda}{16\pi^2 N_e}\left(\dfrac{H}{\Lambda_{\text{sp}}}\right)^3 \;,
\end{equation}
\begin{equation}
    r=\dfrac{8}{\lambda^2 N_e^2}\left[1-N_e\dfrac{3\,\gamma\lambda}{16\pi^2}\left(\dfrac{H}{\Lambda_{\text{sp}}}\right)^3\right] \;.
\end{equation}

\subsubsection*{Inverse hilltop inflation}

Finally, we look at inverse hilltop potentials, which can arise naturally in the context of brane-antibrane inflation~\cite{Burgess:2001fx}, but also in moduli inflation for Type IIB model building~\cite{Bansal:2024uzr}
\begin{equation*}
V(\vp)=V_0\left(1-\frac{2(\frac{\lambda}{q})^{-q}}{q-2}\vp^{2-q}\right)^2,\quad1<\alpha=2-\frac{2}{q}<2,\quad\beta_\alpha=\dfrac{(4q)^{2/q}}{2\lambda ^2} \;,
\end{equation*}
where $q>2$ is a free parameter to control the shape of the inverse hilltop potential. It is easy to show that for different values of $q$ (and thus $\alpha$), this model smoothly interpolates between chaotic ($q\to2$) and Starobinsky ($q\to\infty$) inflation. For this work, we have considered two benchmark cases. The first is $q=4$, corresponding to 
\begin{equation}
    V(\phi) = V_0 \left( 1 - \left(\frac{\lambda}{q}\right)^{-4}\phi^{-2} \right)^{2} \; , 
\end{equation}
which gives
\begin{equation}
    n_s-1\simeq-\dfrac{3/2}{N_e}-\dfrac{4}{\lambda^2 N_e^{3/2}}-\dfrac{3\gamma/\lambda}{8\pi^2}\left(\dfrac{H}{\Lambda_{\text{sp}}}\right)^3 \dfrac{1}{N_e^{3/4} } \;,
\end{equation}
\begin{equation}
    r=\dfrac{32}{\lambda^{2} N_e^{3/2}}\left[1-N_e^{3/4}\dfrac{3\gamma\lambda}{32\pi^2}\left(\dfrac{H}{\Lambda_{\text{sp}}}\right)^3\right] \;,
\end{equation}
and $q-2=2/3$, corresponding to 
\begin{equation}
    V(\phi) = V_0 \left[1- 3  \left(\frac{3 \lambda}{8}\right)^{-8/3} \vp^{-2/3}\right]^2 \; ,
\end{equation}
which gives
\begin{equation}
    n_s-1\simeq-\dfrac{5/4}{N_e}-\dfrac{8 (2/3)^{\frac{3}{4}}
    }{\lambda^2 N_e^{5/4}}-\dfrac{3\gamma/\lambda}{8\pi^2}\left(\dfrac{H}{\Lambda_{\text{sp}}}\right)^3 \dfrac{2 \sqrt{2} \left(\frac{2}{3}\right)^{3/4}}{N_e^{5/8} } \; ,
\end{equation}
\begin{equation}
    r=\dfrac{64 (2/3)^{\frac{3}{4}}
    }{\lambda^2 N_e^{5/4}}\left[1-\dfrac{N_e^{5/8}}{2 \sqrt{2} \left(\frac{2}{3}\right)^{3/4}}\dfrac{3\gamma\lambda}{16\pi^2}\left(\dfrac{H}{\Lambda_{\text{sp}}}\right)^3\right] \; .
\end{equation}
For the analysis in~\cref{sec:pheno}, and in particular, for~\Cref{fig:multiple_constraints} and~\Cref{fig:testfig}, we have fixed $\lambda=4$ and $\lambda = \frac{8\ 10^{3/8}}{3^{5/8}}$, respectively for these two cases.

\bibliographystyle{utphys}
\bibliography{vbf}

\end{document}